\documentclass[journal]{IEEEtran}

\usepackage{amsmath,amsfonts}
\usepackage{graphicx}
\usepackage{booktabs}
\usepackage[normalem]{ulem}
\usepackage{wrapfig}
\usepackage{paralist}
\usepackage{xcolor}
\usepackage{cite}
\usepackage[hidelinks]{hyperref}
\usepackage{cleveref}

\graphicspath{{figs/}{figs/imgs/}{./}}
\def\BibTeX{{\rm B\kern-.05em{\sc i\kern-.025em b}\kern-.08em
    T\kern-.1667em\lower.7ex\hbox{E}\kern-.125emX}}

\begin{document}


\title{AdaLens: Interactive Storyline for Monitoring and Steering Long-Running Agentic Data Analysis}

\author{
Yangtian Liu, 
Yan Miao,
Shuhan Liu, 
Yunfan Zhou, 
Dae Hyun Kim, 
Di Weng,
and Yingcai Wu

\thanks{Y. Liu, Y. Miao, S. Liu, Y. Zhou, and Y. Wu are with the State Key Lab of CAD\&CG, Zhejiang University, Hangzhou, Zhejiang, China.
E-mail: \{yt-liu, ymiao, shliu, yf.zhou, ycwu\}@zju.edu.cn.}
\thanks{D. H. Kim is with the Department of Computer Science and Engineering, Yonsei University, Seoul, Republic of Korea.
E-mail: dhkim16@yonsei.ac.kr.}
\thanks{D. Weng is with the School of Software Technology, Zhejiang University, Ningbo, Zhejiang, China.
E-mail: dweng@zju.edu.cn. 
D. Weng is the corresponding author.}
\thanks{This work has been submitted to the IEEE TVCG for possible publication. }
}

\markboth{Journal of \LaTeX\ Class Files,~Vol.~18, No.~9, September~2020}%
{How to Use the IEEEtran \LaTeX \ Templates}

\maketitle


\begin{abstract}
Large language models are pushing data science toward increasingly autonomous and agentic workflows, with recent systems already supporting multi-step and long-running analyses.
As these workflows become more autonomous, conventional interfaces no longer provide adequate support for two critical requirements: observability for understanding an agent's evolving reasoning and evidence, and steerability for redirecting low-value directions or deepening promising ones during execution.
Existing interactive approaches improve process visibility and open intervention points, but they remain largely designed for discrete, turn-by-turn exchanges rather than the parallel branches and evolving decision structures of long-running agentic analysis.
We study this need as interactive oversight in long-running agentic data analysis and present AdaLens, an interactive system for monitoring and steering ongoing runs.
AdaLens combines a storyline-based representation that unifies analytical plans, execution progress, intermediate findings, and data-column involvement with steering interactions grounded in these analytical elements for directional guidance and execution control.
We evaluate AdaLens through two case studies and a user study, examining how it supports analysts in monitoring and steering long-running agentic data analysis.
\end{abstract}



\begin{figure*}[t]
  \centering
  \includegraphics[width=\textwidth]{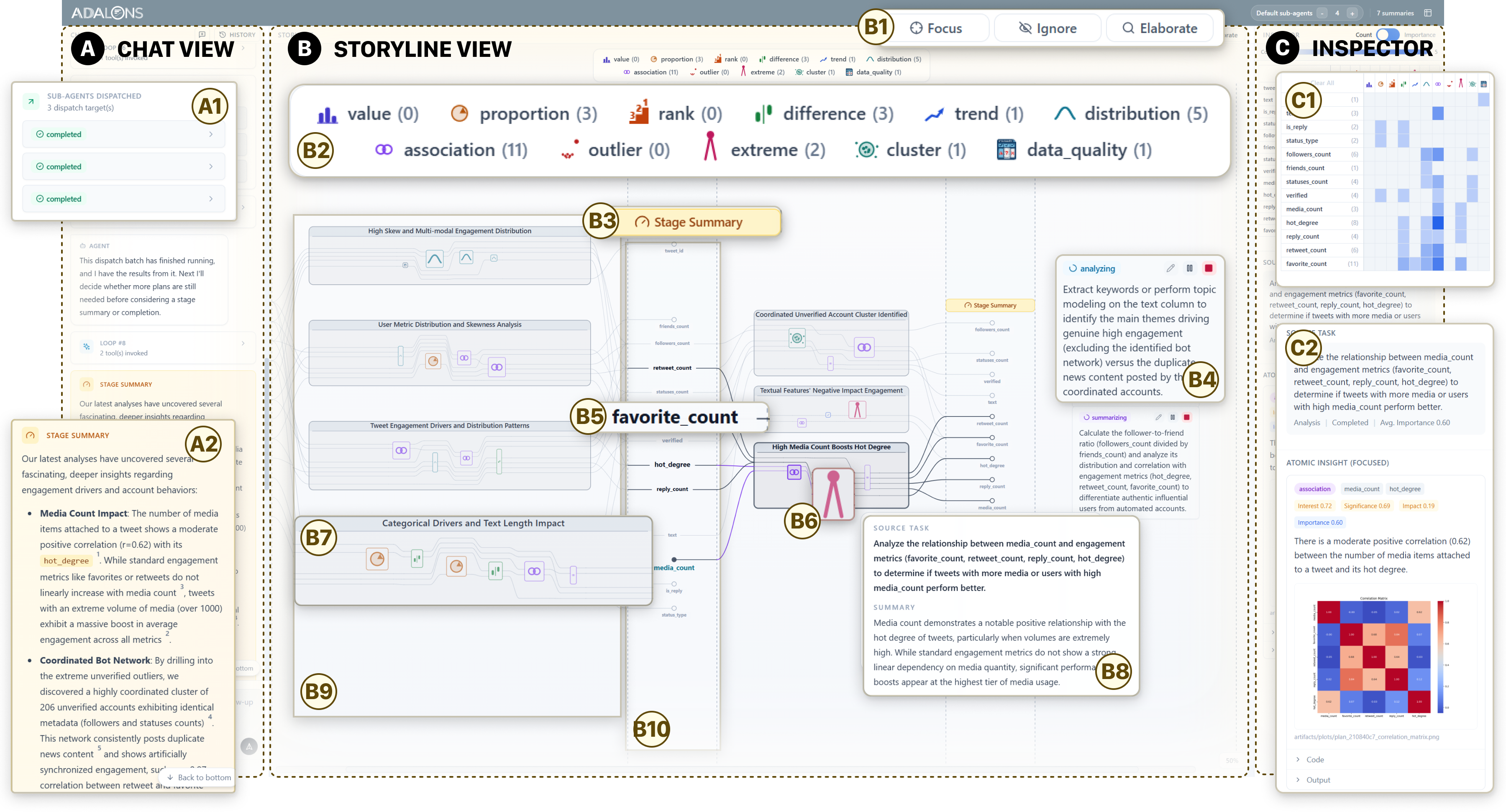}
  \caption{
    AdaLens interface for monitoring and steering long-running agentic data analysis through three coordinated views.
    (A) The chat view records user and system messages, dispatched plans (A1), and stage/final reports (A2).
    (B) The storyline view serves as the primary workspace, combining steering tools (B1), insight-type legends (B2), report anchors (B3), plan/summary cards (B4, B7), data-column traces (B5), atomic insight glyphs (B6), on-demand detail popovers (B8), and the alternation between finding regions (B9) and converge regions (B10) to reveal stepwise progression and data-grounded analytical lineage over time.
    (C) The inspector view summarizes coverage across columns and insight types (C1) and reveals the selected task or finding with linked evidence such as plots, code, and outputs (C2).
  }
  \label{fig:teaser}
\end{figure*}


\section{Introduction}
\IEEEPARstart{L}{arge} language models (LLMs) are driving data science toward increasingly autonomous workflows~\cite{ma2023insightpilot,chen2024lightva}, freeing analysts from tedious manual execution to focus on high-level sensemaking.
Recent agentic systems~\cite{he2025deepanalyze,manatkar2024quis,zhu2025dataagents} push this trend further by enabling LLM-based agents to conduct end-to-end data analysis with minimal human intervention.
Given an analytical objective and datasets, the agent formulates analytical plans, executes them to gather evidence, and iteratively refines or expands its investigation based on intermediate insights.
Consequently, agentic data analysis becomes a multi-step, branching, and long-running process that unfolds through successive cycles of planning, execution, and synthesis over an extended timeframe.
Across these cycles, the agent pursues multiple analytical directions and uses accumulated findings and evidence to determine what to analyze next.
As the agent operates with increasing autonomy, the analyst's role shifts from a step-by-step driver to a high-level overseer, making two complementary interface requirements especially important: \textit{observability} and \textit{steerability}~\cite{liu2024waitgpt,gu2024steering}.
\textit{Observability} concerns whether analysts can maintain a coherent understanding of the agent's evolving analytical process---which directions have been explored, what insights have emerged, what evidence supports the results, and how the run has unfolded as a whole.
Existing approaches, designed around turn-by-turn interaction, enhance visibility by exposing fine-grained details such as transformation provenance~\cite{feng2024xnli} and intermediate data states~\cite{liu2024waitgpt}, or by surfacing high-level key insights~\cite{zhao2025insightlens}. 
However, they restrict analysts to inspecting outcomes sequentially, failing to capture the branching directions and interdependent analytical threads that characterize agentic analysis.
\textit{Steerability} concerns whether analysts can effectively redirect the agent's course when they spot opportunities for improvement---pruning an unproductive branch, prioritizing a promising thread, or deepening a specific finding.
Existing steering mechanisms typically operate at the level of individual analytical units: editing task decompositions or plan steps before execution~\cite{gu2024steering, chen2024lightva}, manipulating nodes in a dataflow graph~\cite{castelo2025flowco}, or surgically correcting individual operations mid-execution~\cite{liu2024waitgpt}.
While effective for controlling specific steps, these mechanisms do not support redirecting the overall course of an evolving investigation where the agent has accumulated context across many interdependent steps.
When broader redirection is needed, analysts must fall back on natural language feedback~\cite{majumder24datavoyaer}, which still forces them to mentally reconstruct this accumulated context to articulate directional intent.
Therefore, our goal is to support analysts in monitoring and steering multi-step, branching, and long-running agentic data analysis through structured representations of the agent's evolving analytical process and interaction mechanisms grounded in these representations.
Grounded in the characteristics of long-running agentic data analysis and informed by the recurring practice challenges identified through our six-month collaboration with three experts, as detailed in \cref{sec:practical-challenges}, we formulate two key design challenges:

\textbf{DC1 (\textit{Observability}): How to build legible representations of evolving analytical processes?}
Long-running agentic data analysis produces a growing, heterogeneous collection of artifacts distributed across many interdependent steps, making the raw process difficult to follow.
Legibility requires addressing two intertwined difficulties.
First, analysts need to inspect the process at \textit{multiple levels of granularity}, from tracking how analytical directions branch to examining specific findings and their supporting evidence.
Because the analytical structure emerges dynamically as the agent iterates, the representation must accommodate new elements without disrupting the analyst's orientation.
Second, the representation must preserve \textit{analytical lineage}: how findings and their data grounding remain connected as the process evolves.
Without this traceability, analysts may struggle to understand how the analysis has evolved and how its findings remain grounded in the data.


\textbf{DC2 (\textit{Steerability}): How to enable precise and timely steering of ongoing analysis?}
Given a legible representation, the second challenge is enabling analysts to act on it effectively.
Steering is difficult because interventions span multiple interdependent levels: analysts may need to redirect the agent's high-level analytical direction while also managing the lifecycle of individual plan threads.
These levels are coupled, as a directional shift may render active threads obsolete, and a newly created plan must be coordinated with the agent's accumulated context.
Beyond choosing what to steer, analysts face the problem of when and how to intervene: the agent's state evolves continuously, making intervention opportunities transient, while free-form natural language lacks the precision to express intent without burdening analysts with context reconstruction.
Effective steering therefore requires interaction mechanisms grounded in the elements of the representation itself, letting analysts express intent directly on the artifacts they are inspecting.

To address these challenges, we present AdaLens (\textbf{A}gentic \textbf{d}ata \textbf{a}nalysis \textbf{Lens}), an interactive system for monitoring and steering long-running agentic data analysis.
For DC1, we design a storyline-based representation that supports multi-granularity inspection by organizing analytical plans, summaries, and atomic insights as visual elements along an evolving timeline, while preserving analytical lineage by threading data columns through the storyline as persistent characters whose participation across steps and findings remains visually traceable.
For DC2, we design steering interactions grounded in these storyline elements, enabling analysts to redirect the agent's analytical direction by acting on summaries, atomic insights, and data columns (Focus, Ignore, Elaborate) and to manage the lifecycle of plan threads through direct execution controls (Create, Launch, Pause, Modify, Terminate).
We evaluate AdaLens through two case studies and a user study.
The case studies illustrate how analysts used the storyline to monitor evolving analyses and apply both intention-level steering and execution-level control; the user study reports a mean System Usability Scale (SUS) score of 87.08 and positive participant assessments of \textit{observability} and \textit{steerability}.

In summary, our core contributions are as follows:
\begin{compactitem}
  \item We formulate the problem of monitoring and steering long-running agentic data analysis, identifying key challenges around building legible representations of evolving analytical processes and enabling precise steering grounded in these representations.
  \item We propose AdaLens, an interactive system that integrates a storyline-based representation supporting multi-granularity inspection and data-grounded analytical lineage with steering interactions that operate at both the intention level and the execution level.
  \item We evaluate AdaLens through two case studies on two real-world datasets and a user study on representative analytical tasks.
\end{compactitem}


\section{Related Work}
We review existing research on empowering data analysis with LLMs, visualizing data analysis processes, and interacting with LLMs.

\subsection{Empowering Data Analysis with LLMs}
Large language models (LLMs) have enabled more open-ended data analysis through broad domain knowledge and strong reasoning capabilities~\cite{zhu2025dataagents}, and the autonomy granted to them has grown progressively.
Early tools focused on generating standalone visualizations from natural language~\cite{vizgpt2024,dibia2023lida} and refining this mapping through prompt scaffolding, few-shot examples, and multimodal interactions~\cite{chen2025prompt4vis,liu2025interchat}, though effective use still depends on prior analytical skill and careful inspection of intermediate outputs~\cite{gu2024analysts,rao2025noteveryone}.
Subsequent efforts extended LLM involvement to progressive workflows~\cite{ma2023insightpilot,chen2024lightva,li2025jupybara} that externalize analytical context~\cite{wang2025noteex,wang2025formulator2} and support iterative exploration, report adaptation, and hypothesis refinement~\cite{castelo2025flowco,epperson2025respark,kang2025guardrails}.
These systems shift the LLM's role from producing standalone artifacts to coordinating an evolving analytical process, though multi-step reasoning and intent disambiguation remain challenging~\cite{li2024tapilot,zhang2025condabench}.
More recently, autonomous agents integrate planning, tool use, and iterative refinement for end-to-end workflows~\cite{hong2025datainterpreter,agenticdata2025,he2025deepanalyze,manatkar2024quis}, with some extending to automated report writing and narrative generation~\cite{wang2025dagent,he2024datanarrative}.
However, benchmarks suggest that reliably achieving fully autonomous analysis remains an open challenge~\cite{hu2024infiagent,siddiqui2025insightbench,zhang2025dacomp,zhang2026dsaeval}.
Building on this line of work, AdaLens employs an orchestrator--worker architecture to support long-running agentic data analysis, enabling autonomous and multi-step reasoning.

\subsection{Visualizing Data Analysis Processes}
Visualizing the data analysis process helps users trace and revisit how results were reached~\cite{ragan2015provenance,xu2019provenance,gathani2024provenance}.
We review such visual representations at different abstraction levels.

At the data transformation level, systems make intermediate states visible through live profiling~\cite{epperson2023autoprofiler}, animated pipeline explanations~\cite{pu2021datamations}, or inspectable query decompositions~\cite{deutch2021diy,feng2024xnli}.
For LLM-driven analysis, WaitGPT~\cite{liu2024waitgpt} renders streaming code as real-time dataflow graphs with table glyphs encoding data shape changes, while ViseGPT~\cite{chen2025visegpt} visualizes test results as Gantt charts for error localization.

At the workflow level, flow graphs foreground dependencies among data states and operations.
NoteFlow~\cite{epperson2025noteflow} and NoteEx~\cite{wang2025noteex} extract or let analysts define inter-cell data dependencies for bidirectional tracing in computational notebooks, while Flowco~\cite{castelo2025flowco} and LightVA~\cite{chen2024lightva} use dataflow and task-flow graphs respectively to structure the authoring and planning of analysis pipelines.
Tree views externalize branching exploration. Data Formulator 2~\cite{wang2025formulator2} organizes iterative visualization authoring into Data Threads for branching and context reuse, Kang et al.~\cite{kang2025guardrails} structure hypothesis exploration with ordered node-link diagrams, and Narrative Scaffolding~\cite{sultanum2026narrative} maintains branching narrative trees that record reasoning evolution.
Timeline and history views, such as those combining artifact and search interfaces for fragment-level version foraging~\cite{head2019foraging} or recording sensemaking provenance for post-hoc analysis~\cite{nguyen2016sensepath}, help reconstruct how a result was reached.

At a higher abstraction, systems move beyond process steps to organize findings and track exploration coverage: InsightLens~\cite{zhao2025insightlens} clusters insights hierarchically with a temporal minimap, while Snowy~\cite{srinivasan2021snowy} visualizes attribute and intent coverage to guide exploration breadth.

AdaLens frames long-running agentic data analysis as an evolving narrative and adopts a storyline-based visualization to jointly support temporal progression, multi-granularity inspection, and data-grounded analytical lineage.

\subsection{Interacting with LLMs}
Researchers have advanced interaction techniques that give users more control over LLMs, moving beyond standard chatbot interfaces.

A core challenge is the ``gulf of envisioning''~\cite{subramonyam2024bridging}: users struggle to anticipate how prompts translate into outputs, which imposes substantial metacognitive demands~\cite{tankelevitch2024metacognitive} and motivates tighter alignment between user intent and model behavior~\cite{shen2024bidirectional}.
To close this gap, one line of work supports post-hoc inspection, including provenance tracking~\cite{kim2024hallmark}, explanatory diagnostics~\cite{feng2024xnli}, and empirical studies of analyst verification behavior~\cite{gu2024analysts, kim2024analysts}, while another opens pre-output intervention points such as editable task decompositions~\cite{gu2024steering} and externalized intent-to-task mappings~\cite{jiang2025neurosync}.
For multi-agent and code-generation workflows, recent tools provide checkpoint-based counterfactual replay~\cite{cheng2025agdebugger}, real-time visual monitoring of streamed code~\cite{liu2024waitgpt}, and constraint-based test generation for automated error localization~\cite{chen2025visegpt}.

Another stream of research moves beyond single-text prompting by externalizing plans, context, and semantic structures as manipulable objects.
Several systems externalize intermediate plans and processing steps as objects that users can inspect and edit~\cite{wu2022aichains, cai2024lowcodellm, zhang2024cocoa, gu2025magneticui, cai2024coladder}.
Spatial interfaces move beyond linear dialogue by introducing navigable structures such as interactive node-link graphs~\cite{jiang2023graphologue}, multilevel abstraction hierarchies~\cite{suh2023sensecape}, dimension-driven design-space maps~\cite{suh2024luminate}, freeform canvases for side-by-side comparison~\cite{wu2024intelligentcanvas}, and editable conversational memory~\cite{lee2023memorysandbox}.
Multimodal interaction further reduces prompt ambiguity by combining natural language with direct manipulation, sketching, or dynamically synthesized widgets~\cite{masson2024directgpt, vaithilingam2024dynavis, liu2025interchat, wu2025vispilot, chen2024sketchthengenerate}.
Beyond richer input modalities, ProactiveVA~\cite{chen2025proactiveva} shifts the locus of initiative by letting agents detect user difficulties and offer context-aware assistance unprompted.
These interaction patterns, spanning structured plans, spatial workspaces, and multimodal or proactive support, have been adopted in iterative data analysis~\cite{wang2024formulator, wang2025formulator2, wang2025noteex, li2025jupybara, epperson2025respark, chen2024lightva}, data-driven sensemaking and storytelling~\cite{sultanum2026narrative, wang2025inreactable, kang2025guardrails}, and creative authoring interfaces~\cite{chung2022talebrush, angert2023spellburst, pan2024smartboard, qian2024shapeit}.

Unlike prior approaches designed around turn-by-turn interaction and discrete checkpoints, AdaLens supports intervention within continuous agentic data analysis through direct manipulation interactions grounded in visible analytical elements, enabling analysts to steer the evolving analytical process asynchronously and in situ.

\section{Informing the Design}
\label{sec:informing-design}
AdaLens is designed for analysts who already use LLM-based assistants or agentic coding tools in their data analysis workflows.
In these workflows, agents can carry out multi-step investigations beyond a single prompt--response exchange, making it important for analysts to follow intermediate progress and intervene when useful.
To ground the design of AdaLens in realistic analytical practice, we collaborated closely with three researchers who actively use such tools in their data analysis work.
We recruited them through direct verbal or instant-messaging invitations and provided no compensation.\footnote{All experts in the formative collaboration and case studies, as well as all user-study participants, provided informed consent before taking part.}
EA is an assistant professor whose research focuses on emerging data analytical techniques driven by agentic LLMs.
EB and EC are Ph.D. candidates in time-series and urban data analysis, respectively, who both frequently employ agentic LLM tools in processing relevant data.
Together, they bring both research perspective and hands-on experience using LLM-based assistants and agentic coding tools for data analysis.
The tools they currently use (e.g., ChatGPT~\cite{openai2022chatgpt}, Julius AI~\cite{juliusai}) already support extended agentic analysis, but mainly through natural-language interaction and sequential textual or code-oriented outputs, motivating our focus on structured visual oversight to address shared workflow needs in agentic data analysis.

Our collaboration broadly followed the design study methodology of Sedlmair et al.~\cite{DBLP:journals/tvcg/SedlmairMM12}.
Over approximately six months, we met with the experts in weekly hour-long sessions spanning the core phases of the methodology.
In the \emph{discover} phase, we asked the experts to walk through concrete analysis scenarios, describe how they would inspect and intervene in an ongoing agent-driven analysis, and identify what information they would need to do so effectively.
These discussions led to the problem formulation, current practice challenges, and design goals described below.
In the \emph{design} and \emph{implement} phases, we translated these goals into prototype designs and continued meeting with the experts to iteratively evaluate and refine the visual encodings, interaction mechanisms, and system workflow until the experts confirmed that the design adequately supported the identified design goals.

\subsection{Problem Formulation}
\label{sec:problem-formulation}
Drawing on our discussions with the experts, we formulate \emph{agentic data analysis} as a process in which an agent receives a high-level analytical goal and a dataset, then autonomously advances the investigation through repeated cycles of planning, execution, and synthesis.
Unlike turn-by-turn LLM assistance, its central unit is not a single conversational turn but an evolving \emph{run} that persists beyond any single interaction, during which the agent can pursue several candidate directions concurrently and revise them as evidence accumulates.
Overseeing such a process requires analysts to understand its evolving structure and intervene as it unfolds---a problem we term \emph{interactive oversight}.
We characterize a run through the following recurring analytical elements and their relationships:

\begin{compactitem}
\item \textbf{Step.} One round in which the system updates the run state by deciding what to analyze next, executing the corresponding analysis, and producing the new results.
\item \textbf{Plan and plan thread.} A \emph{plan} is an explicit analytical intention for one direction of inquiry. A \emph{plan thread} refers to the ongoing lifecycle of that direction as it progresses from pending through active execution to completion, and may be paused or terminated by the analyst along the way.
\item \textbf{Atomic insight.} A fine-grained finding that states a specific observation about the data, such as a trend, an outlier, or a correlation, linked to concrete supporting evidence such as code, plots, and textual outputs.
\item \textbf{Summary.} When a thread completes its assigned analysis, the system produces a summary that synthesizes the resulting atomic insights into a thread-level conclusion.
\item \textbf{Data column.} The specific parts of the dataset implicated in a plan or finding. Data columns recur across multiple threads and steps, providing a persistent connective medium that grounds the analysis in the underlying data.
\end{compactitem}

These elements do not exist in isolation.
We use the term \emph{analytical lineage} to refer to the web of relationships that connects them: plans initiate threads, completed threads yield atomic insights that are synthesized into summaries, and those findings can motivate, refine, or constrain later plans in subsequent steps.
This lineage is inherently difficult to follow because threads run concurrently within each step, data columns split across those threads and reconverge in later steps, and each completed thread produces a nested hierarchy of summaries and insights with their own internal structure.
The underlying data unifies this lineage: every element is grounded in specific data columns, plans target them, insights are derived from them, and threads operating on overlapping columns share observable data grounding and are analytically related through their common data involvement even when they belong to different steps.
Shared data therefore provides a natural connective tissue through which analytical lineage can be traced and made visible.

\subsection{Challenges in Current Practice}
\label{sec:practical-challenges}
Across our ongoing collaboration with the experts, we identified three recurring practice challenges in their experiences with long-running agentic data analysis.
Specifically, PC1 and PC2 characterize the observability challenge (DC1), whereas PC3 characterizes the steerability challenge (DC2).

\textbf{PC1: Analysts struggle to inspect the accumulating analytical artifacts of a run.}
As a run progresses, it continuously produces analytical artifacts at multiple levels of granularity, ranging from run-level steps and analytical plans to fine-grained findings and supporting evidence.
Current tools often do not organize these artifacts for structured inspection, leaving them difficult to access, locate, or inspect.
Analysts may therefore need to locate and review relevant information across extensive intermediate outputs, making it difficult to examine specific findings without losing sight of the run's overall state.
For example, EB described intermediate conclusions as difficult to inspect and follow up on in context, while EC reported having to explicitly save scripts and intermediate artifacts that would otherwise remain temporary and difficult to manage or inspect.
EA further noted that existing end-to-end tools often prioritize final reports while providing little structured access to the underlying analytical process, making intermediate steps cumbersome to trace.

\textbf{PC2: Analysts struggle to trace the evolving analytical structure of a run.}
A run can unfold through multiple analytical directions that proceed concurrently and extend across successive steps.
Although intermediate outputs collectively document the activities and findings produced throughout the run, the relationships among these analytical artifacts and their grounding in the underlying data remain implicit.
Consequently, analysts find it difficult to reconstruct the analytical structure of the run and understand how it evolved as a connected process.
For example, EC used a separate collaborative document to record analytical plans and track progress.
However, EC noted that loose and sometimes outdated links among the document, intermediate artifacts, and results could obscure the overall analytical trajectory.
EB likewise found it difficult to connect intermediate conclusions to the information from which they were derived, as the conclusions were numerous, inconsistently presented, and sometimes embedded in the agent's intermediate reasoning.

\textbf{PC3: Analysts struggle to intervene effectively in a run as it unfolds.}
As a run unfolds, analysts may identify promising directions to pursue, irrelevant directions to deprioritize, unproductive plans whose execution should be adjusted, or new analytical ideas worth exploring.
In current practice, analysts largely rely on natural-language instructions to express such intentions, requiring them to translate their judgments and relevant analytical context into actionable requests.
Consequently, they find it difficult to intervene efficiently and precisely, both in redirecting future investigation and in regulating the execution of ongoing plan threads.
For example, the tools used by EC did not support pausing an ongoing execution, and modifying or terminating one task required interrupting the overall analytical process.
EB reported that steering a direction that emerged during a run required repeated follow-up questions to reconstruct its context, and EA similarly noted that current agentic tools offered limited support for redirecting the analysis when an agent began to drift from the high-level analytical goal.

\subsection{Design Goals}
\label{sec:design-goals}
From the practice challenges above, we derive four design goals for interactive oversight in long-running agentic data analysis.
G1 and G2 address the observability challenges identified in PC1 and PC2, respectively, while G3 and G4 address the intervention challenge described in PC3 at the levels of analytical direction and ongoing execution.

\textbf{G1: Represent analytical elements across multiple levels of granularity.}
The system should expose the key elements of a run, from active plan threads and step-level progress to summaries, atomic insights, supporting evidence, and involved data columns.
Such a multi-granular representation is necessary because analysts overseeing long-running analyses need to alternate between global orientation and close inspection of specific findings.
It should support fluid navigation across these levels so that analysts can shift between monitoring overall progress and examining individual results.

\textbf{G2: Visualize data-grounded analytical lineage across plans, summaries, and atomic insights.}
The system should make visible how plans, summaries, and atomic insights relate to one another through the data columns they share.
Since overlapping data columns connect elements both within and across steps, they provide a structural basis for tracing how the process evolves, how threads diverge or converge, and how findings share data grounding.
This goes beyond exposing isolated artifacts: analysts should be able to follow the data-grounded continuity of the run rather than manually reconstruct it from scattered outputs.

\textbf{G3: Steer analytical direction by interacting with visible elements.}
The system should allow analysts to influence what the agent investigates next by acting on the summaries, atomic insights, and data columns they observe, signaling which directions to prioritize, suppress, or deepen.
Such steering should be grounded in visible elements of the ongoing run, enabling analysts to express intentions precisely and in context.
This reduces the burden of reconstructing and restating accumulated context through free-form natural language.

\textbf{G4: Regulate the lifecycle of ongoing plan threads.}
The system should allow analysts to create, launch, pause, modify, and terminate plan threads as the run evolves.
This control is necessary because the value of a thread can change over time, and analysts need to coordinate computational effort with emerging findings and domain knowledge.
These actions should take effect during the run without requiring analysts to interrupt the overall analytical process.

\section{AdaLens}
\label{sec:system}
This section first presents the orchestrator--worker framework that drives AdaLens (\cref{sec:system-framework}).
It then describes how the coordinated views, particularly the storyline visualization, help analysts understand the ongoing analytical process (\textbf{DC1}; \cref{sec:system-understanding}) and how the interaction mechanisms enable analysts to steer and control the run (\textbf{DC2}; \cref{sec:system-steering}).
Finally, it provides implementation details (\cref{sec:system-implementation}).

\subsection{System Framework}
\label{sec:system-framework}
\begin{figure*}[t]
  \centering
  \includegraphics[width=\textwidth]{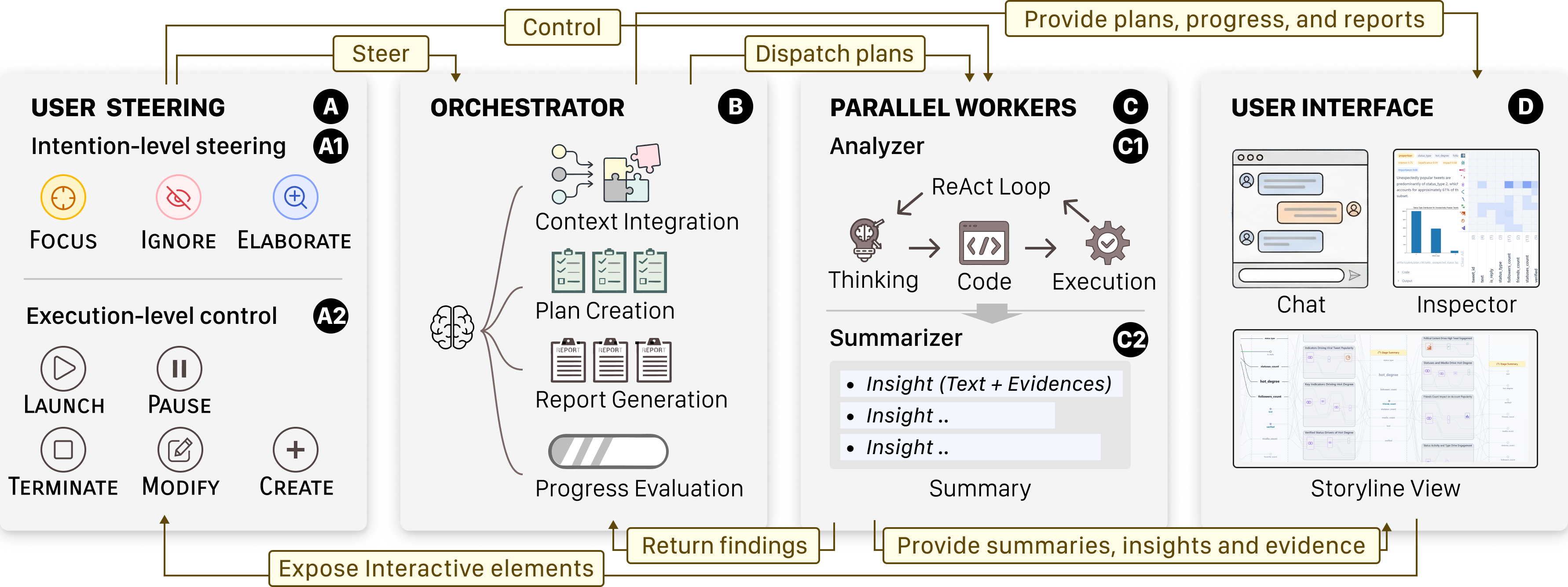}
  \caption{AdaLens framework for interactive oversight in long-running agentic data analysis.
  Analysts steer the run through intention-level actions and execution-level controls (A). 
  The orchestrator manages context integration, plan creation, report generation, and progress evaluation (B).
  The orchestrator dispatches plans to parallel workers whose analyzer and summarizer produce summaries and atomic insights with supporting evidence (C), and these intermediate analytical elements are surfaced through the chat, inspector, and storyline views (D).}
  \label{fig:framework}
\end{figure*}

Building on the process model in \cref{sec:problem-formulation}, AdaLens instantiates agentic data analysis as an autonomous and multi-step process while preserving observability and steerability for analysts throughout the run. 
To drive the autonomous workflow, AdaLens adopts an orchestrator--worker architecture consisting of an orchestrator agent and multiple worker agents.
The orchestrator handles high-level reasoning and orchestrates the overall analytical process (\cref{fig:framework}B).
At each analytical step, the orchestrator decides how the workflow should proceed, including whether to generate new plans, dispatch them to workers, evaluate the progress, and synthesize findings.
Each worker is responsible for executing a specific analytical plan issued by the orchestrator (\cref{fig:framework}C).
Internally, it comprises two sequential components: an analyzer (\cref{fig:framework}C1) and a summarizer (\cref{fig:framework}C2).
The analyzer performs iterative reasoning and analysis under a ReAct-style prompting paradigm~\cite{yao2023react}, including generating code, executing it, interpreting intermediate results, and reflecting on the current progress.
The summarizer then consolidates the analysis process and outcomes into a structured finding hierarchy, including a summary and a set of atomic insights, each linked to concrete supporting evidence (e.g., code, plots, and textual outputs).
These findings are returned to the orchestrator and utilized as context for subsequent decision-making.

Complementing this autonomy, AdaLens is designed to keep analysts informed and in control throughout the long-running analysis.
To support observability, the system preserves essential analytical artifacts produced during execution, including plans, summaries, atomic insights, and their supporting evidence, making them available for inspection on the user interface (\cref{fig:framework}D) at any point.
To support steerability (\cref{fig:framework}A), the system captures analyst interventions in real time at both the intention level for the orchestrator (\cref{fig:framework}A1) and the execution level for workers (\cref{fig:framework}A2), enabling the analysis to adjust its trajectory without terminating and restarting the run.
These two mechanisms form the architectural basis for the interactive oversight features detailed in the following sections.

\subsection{Understanding the Ongoing Analytical Process}
\label{sec:system-understanding}
To maintain the observability of the ongoing analytical process for analysts (\textbf{G1, G2}), AdaLens organizes the analytical elements introduced in \cref{sec:problem-formulation} into three coordinated views: the storyline view (\cref{fig:teaser}B), the chat view (\cref{fig:teaser}A), and the inspector view (\cref{fig:teaser}C).
They reveal how the analysis is structured and proceeds, how findings accumulate over time, and what evidence supports each result.

\subsubsection{Storyline View}
Inspired by storyline visualization techniques~\cite{tanahashi2012design,tang2019istoryline}, we conceptualize long-running agentic data analysis as an evolving narrative. 
To elucidate this dynamic progression, the storyline view integrates diverse analytical elements into a unified temporal representation, serving as AdaLens's primary interface. 
We detail its visual elements, layout design, design rationale, and inspection-oriented interactions below.

\textbf{Visual elements.}
The storyline view visualizes three types of key analytical elements (\cref{sec:problem-formulation}) at multiple levels of granularity (\textbf{G1}).
\underline{\textit{(1) Plans and summaries.}}
Plan threads dynamically appear as analytical cards (\cref{fig:teaser}B4) showing the plan description and execution status (e.g., pending, analyzing, summarizing, completed, terminated, or failed). 
Once a thread completes, it is converted into a summary card (\cref{fig:teaser}B7) displaying a brief label and extracted atomic insights.
These cards form the backbone of the storyline, marking key analytical events and conclusions as the analysis progresses.
\underline{\textit{(2) Atomic insights.}}
Within each summary card, AdaLens exposes fine-grained findings as atomic insight glyphs (\cref{fig:teaser}B6) 
arranged in order of generation.
Glyph size reflects an importance score computed from statistical indicators (impact and significance) combined with LLM-judged interestingness.
Insight types are informed by~\cite{zhao2025insightlens, wang2020datashot} and encoded with distinct glyphs (\cref{fig:teaser}B2).
This design helps analysts quickly identify important insights and navigate between high-level summaries and detailed findings without leaving the storyline view.
\underline{\textit{(3) Data columns.}}
To visualize recurring data grounding, we render data columns (\cref{fig:teaser}B5) as characters in the narrative of a long-running agentic data analysis.
The label size of each column character reflects the number of atomic insights involving that column in the corresponding analytical step, visually conveying the extent of its involvement in that step.
At the beginning of each step, these characters emerge \raisebox{-0.18em}{\includegraphics[height=1.1em]{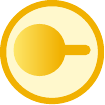}}
 and split \raisebox{-0.18em}{\includegraphics[height=1.1em]{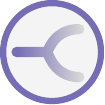}}
 into multiple instances that join different analytical events.
During execution, these column instances visually converge \raisebox{-0.18em}{\includegraphics[height=1.1em]{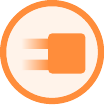}}
 into the summaries and insights in which they are involved.
At the end of each step, they merge back \raisebox{-0.18em}{\includegraphics[height=1.1em]{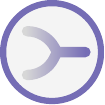}}
 into their original columns and either exit \raisebox{-0.18em}{\includegraphics[height=1.1em]{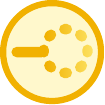}}
 or continue \raisebox{-0.18em}{\includegraphics[height=1.1em]{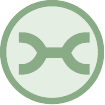}}
 into the subsequent step, depending on whether they remain involved in the next step.
Through this metaphor, analysts can naturally follow how data columns dynamically participate across the analytical process and how findings remain grounded in the data.

In addition to these core elements, the storyline view includes report anchors (\cref{fig:teaser}B3), which are compact markers for the orchestrator's stage-level or final textual syntheses.
Each anchor in the storyline links to the corresponding detailed report in the chat view.

\textbf{Layout design.}
To make the analytical lineage defined in \cref{sec:problem-formulation} legible within a single temporal workspace (\textbf{G2}), the layout arranges the storyline elements and connects them through column-character trajectories while maintaining visual clarity, compactness, and stepwise progression.
It follows a timeline-based storyline metaphor in which analytical steps are arranged from left to right and the view alternates between \textit{finding regions} (\cref{fig:teaser}B9) and \textit{converge regions} (\cref{fig:teaser}B10).
A finding region corresponds to one analytical step and contains summary cards with their nested atomic-insight glyphs, whereas a converge region lies between adjacent finding regions and gathers the column threads that continue from one step to the next.
This alternation makes visible how recurring data columns are distributed across findings within a step and either exit or continue into subsequent steps.

These spatial relations have specific meanings.
Horizontal placement encodes stepwise progression; cards grouped within a finding region belong to the same analytical step, and atomic-insight glyphs nested within a card belong to that summary.
A column trajectory entering a card or glyph indicates that the corresponding artifact involves that column.
Within a converge region, instances of the same column merge into a shared lane when the column continues across adjacent steps, then split toward the artifacts that involve it in the next finding region.
The vertical order of these lanes is inherited from the preceding converge region and adjusted through ordering, alignment, and compaction to preserve trace continuity and reduce crossings and wiggles.
Thus, vertical proximity mainly reflects the inherited ordering and compaction of column trajectories, rather than an additional encoding of similarity or relationship strength.

Our optimization principle follows StoryFlow~\cite{liu2013storyflow}: ordering, alignment, and compaction are used to reduce crossings and wiggles while keeping the layout compact.
Unlike classical storyline layouts, however, our setting evolves incrementally, may branch into concurrent analytical threads, and contains nested finding structure rather than a flat set of entities.
We therefore apply these operations progressively from left to right instead of recomputing the entire storyline offline.
In this adaptation, the finding/converge-region structure and the treatment of each summary card as a local storyline are design choices for representing incremental agentic runs.
The algorithmic component is the repeated local application of ordering, alignment, and compaction, retaining the classical objective of preserving trace continuity, reducing crossings and wiggles, and keeping the visualization clear and compact.
When a new finding region appears, it inherits the solved column order from the left converge region to arrange summary cards and their incoming boundary threads; within each summary card, atomic insights are treated as a local storyline whose column threads are again ordered, aligned across adjacent insight positions, and compacted under spacing constraints.
When a new converge region is created, threads belonging to the same column are merged into a shared converge lane; when the next finding region appears, those lanes split again toward different summary cards and further converge onto the relevant insight glyphs.
By reusing the same StoryFlow-inspired principles locally at each transition, the layout remains stable while accommodating newly arrived summaries, atomic insights, and columns.
Additional implementation details of the progressive layout algorithm are deferred to the supplementary material.

\textbf{Design rationale.}
During the design process, we considered both topology-oriented and chronology-oriented representations for organizing an evolving run.
Topology-oriented representations, such as tree, node-link, and dataflow views, can convey ordered progression, branching, and dependencies, but generally assume an explicit topology among entities.
An analytical run instead combines successive steps, concurrent plan threads, containment relationships between summaries and atomic insights, and the recurring involvement of data columns without conforming to a single tree, dependency graph, or dataflow structure.
Encoding these heterogeneous relationships uniformly as nodes and edges could therefore impose structural semantics that are not consistently present.
Chronology-oriented representations, such as Gantt-style timelines and card-based history views, can effectively organize stepwise progression, execution status, and accumulated analytical artifacts, but typically retain data columns as attributes of individual artifacts or expose their recurrence through additional links.
This local encoding leaves analysts to reconstruct how the same data dimensions recur across concurrent threads, findings, and successive steps.
We therefore selected a storyline representation that organizes plans and summaries as analytical events, embeds atomic insights as nested findings, and represents data columns as persistent characters connecting the events in which they participate.
In this way, the storyline preserves stepwise progression while making the data-grounded continuity across otherwise distributed analytical artifacts directly visible.

\textbf{Inspection-oriented interactions.}
The storyline view further supports understanding through a set of inspection-oriented interactions.
Analysts can directly select plan cards, summary cards, atomic insights, and report anchors to activate the corresponding inspection state, with the inspector and chat views synchronized to reveal the associated detailed information.
The selected elements will be highlighted to help analysts maintain their focus.
Hovering over an atomic insight glyph or a summary card displays a popover (\cref{fig:teaser}B8) with detailed descriptions on the fly, allowing analysts to quickly scan and understand key findings without leaving the storyline view.
The view also supports highlighting findings by selecting data columns and insight types of interest.
In addition, the view supports horizontal panning and zooming, helping users navigate large storylines flexibly.

\subsubsection{Chat View}
The chat view (\cref{fig:teaser}A) presents an analysis run as a conversation list that records executed actions, agent responses, and generated reports.
In addition to the initial analytical goal and subsequent user messages, it displays a sequence of system-generated entries corresponding to distinct stages of the workflow, including orchestrator thinking, plan creation, plan dispatch, progress evaluation, finding synthesis, and report generation.
For dispatched plans, the view embeds compact plan cards (\cref{fig:teaser}A1) showing each plan's current status and any corresponding summary produced from that plan.
Stage-level and final reports (\cref{fig:teaser}A2) further include inline references to the corresponding summary cards or atomic insights in the storyline view, allowing analysts to move from textual reports to their associated visual context.
The panel header also provides access to prior conversations, allowing users to switch between runs while preserving the current run context.
Within a run, the chat view is coordinated with the other views: plan cards and summary entries can be selected from the conversation history, and items shown elsewhere in the interface can be traced back to their textual records.
As a result, the chat view functions not merely as a transcript, but as a chronological record of how the analytical process progresses through planning, execution, and synthesis.

\subsubsection{Inspector View}
The inspector view (\cref{fig:teaser}C) enables analysts to move from process-level awareness to detailed examination of intermediate outputs through the coverage grid and the detailed view.

The coverage grid (\cref{fig:teaser}C1) summarizes how extracted atomic insights are distributed across insight types and dataset columns, with color intensity encoding the degree of coverage.
The detailed view (\cref{fig:teaser}C2) supports multiple inspection modes depending on the current selection.
Selecting a plan activates a plan-focused mode, which presents the plan description, current execution status, and an analysis stream composed of live or recovered artifacts such as generated code, textual outputs, error messages, reflections, and plots.
This mode allows analysts to inspect how an individual analytical task was carried out, including partially completed or failed executions.
Selecting a summary or atomic insight activates a finding-focused mode.
In this mode, the inspector presents the source task, the summary text, and the associated atomic insights.
Each atomic insight is shown together with its taxonomy type, related data columns, importance-related metrics, and linked evidence, including a plot preview and expandable code and output panels.
In addition, analysts can select data columns, insight types, or individual cells in the coverage grid to filter and inspect the corresponding atomic insights, which can be further prioritized by their importance.
Together, these coordinated inspection modes help analysts trace high-level findings back to the concrete artifacts that support them.

\subsection{Steering the Ongoing Analytical Process}
\label{sec:system-steering}
To help analysts steer the ongoing analysis (\textbf{G3, G4}), AdaLens enables direct interaction with meaningful analytical elements in the interface, supporting both intention-level steering of the analytical direction and execution-level control of ongoing plan threads.

\subsubsection{Intention-Level Steering via Visual Elements}
To operationalize \textbf{G3}, AdaLens supports three recurrent intention-level steering actions (\cref{fig:teaser}B1): \textsc{Focus}, \textsc{Ignore}, and \textsc{Elaborate}. 
Rather than relying solely on free-form prompting, analysts issue these actions directly from visual elements in the storyline view, and AdaLens incorporates the resulting steering requests together with information from the selected elements into the orchestrator's context at the next analytical step.
Within this mechanism, \textsc{Focus} and \textsc{Ignore} redirect subsequent planning around selected summaries, insights, or columns, while \textsc{Elaborate} requests a narrower follow-up on a selected summary or insight.

\vspace{1mm}
\begin{wrapfigure}[2]{l}{0.3cm}
\vskip-\intextsep
\includegraphics[width=0.5cm]{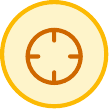}
\vskip-\intextsep
\end{wrapfigure}

\noindent
\textbf{\textsc{Focus}.}
Analysts use \textsc{Focus} when they find one or more directions particularly promising and want the system to continue investing effort in those directions.
For example, after noticing salient keywords in a finding or deciding to further examine specific data columns, analysts may want the system to prioritize those directions in subsequent analysis.
To issue a focus action, the analyst first activates the \textsc{Focus} pen and then clicks a summary card, an atomic insight glyph, or one or more column characters.
For summary cards and atomic insight glyphs, the interaction opens a popover that provides topic keywords distilled from corresponding text, an editable preview of the steering request, and the relevant summary or atomic text as optional background context.
For column characters, the analyst can stage and toggle multiple selected columns before confirmation.
After confirmation, AdaLens records a structured focus request and biases subsequent planning toward devoting more effort to the selected topics, while allowing already running plan threads to continue.

\vspace{1mm}
\begin{wrapfigure}[2]{l}{0.3cm}
\vskip-\intextsep
\includegraphics[width=0.5cm]{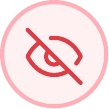}
\vskip-\intextsep
\end{wrapfigure}

\noindent
\textbf{\textsc{Ignore}.}
Analysts use \textsc{Ignore} when a direction appears redundant, low-value, or misaligned with the evolving analytical goal. 
For example, if the system keeps revisiting the same low-yield columns or repeatedly expands a branch that the analyst judges unhelpful, the analyst can suppress that direction before additional time and tokens are spent on it. 
The interaction largely mirrors \textsc{Focus}. 
The analyst activates the \textsc{Ignore} pen and selects a summary card, an atomic insight glyph, or one or more column characters.
Summary- and glyph-level selections open the same keyword- and preview-based popover, while column-level selections support staged multi-column selection before confirmation.
After confirmation, the resulting steering message tells the orchestrator to avoid future investigation along those directions unless they later become necessary for answering the main goal. Existing running plan threads are not cancelled, but subsequent planning shifts away from the ignored target.

\vspace{1mm}
\begin{wrapfigure}[2]{l}{0.3cm}
\vskip-\intextsep
\includegraphics[width=0.5cm]{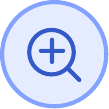}
\vskip-\intextsep
\end{wrapfigure}

\noindent
\textbf{\textsc{Elaborate}.}
Analysts use \textsc{Elaborate} when a finding is important but still underspecified, and they want the system to explain it in greater depth rather than branch broadly. 
For example, after seeing an unusual atomic insight or a summary whose meaning is still unclear, an analyst may want the system to further unpack what drives that result and why it occurs. 
To do so, the analyst activates the \textsc{Elaborate} pen and clicks either a summary card or an atomic insight glyph; unlike \textsc{Focus} and \textsc{Ignore}, \textsc{Elaborate} does not target columns. 
The interaction opens a compact confirmation popover centered on the selected finding, without the keyword chooser used by the other two actions. 
After confirmation, AdaLens issues a narrowly scoped elaboration request that encourages the orchestrator to continue investigating the explanation, mechanism, and root causes of that specific insight while avoiding unnecessary branching into unrelated new plans.

\subsubsection{Execution-Level Control by Regulating Plan Threads}
To operationalize \textbf{G4}, AdaLens supports a set of execution-level control actions spanning the plan lifecycle, including creating, launching, pausing, modifying, and terminating plan threads.
AdaLens supports these actions directly on plan cards, giving analysts fine-grained control over the execution of analytical tasks.

\vspace{1mm}
\begin{wrapfigure}[2]{l}{0.3cm}
\vskip-\intextsep
\includegraphics[width=0.5cm]{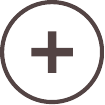}
\vskip-\intextsep
\end{wrapfigure}

\noindent
\textbf{\textsc{Create}} adds a user-authored plan thread for a direction not covered by the current set.
The analyst right-clicks a blank area in the storyline view and enters a new plan in the popover.
If a dispatch is already in progress, the new plan joins the current batch; otherwise, it is dispatched at the next analytical step.

\vspace{1mm}
\begin{wrapfigure}[2]{l}{0.3cm}
\vskip-\intextsep
\includegraphics[width=0.5cm]{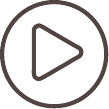}
\vskip-\intextsep
\end{wrapfigure}

\noindent
\textbf{\textsc{Launch}} starts a pending plan thread or resumes a paused one from its preserved phase immediately, rather than waiting for later scheduling.

\vspace{1mm}
\begin{wrapfigure}[2]{l}{0.3cm}
\vskip-\intextsep
\includegraphics[width=0.5cm]{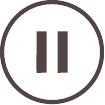}
\vskip-\intextsep
\end{wrapfigure}

\noindent
\textbf{\textsc{Pause}} suspends a running thread that is not urgent enough to continue consuming time and tokens but may be worth revisiting later.
AdaLens preserves the thread's current execution phase and removes it from active execution until the analyst explicitly resumes it.

\vspace{1mm}
\begin{wrapfigure}[2]{l}{0.3cm}
\vskip-\intextsep
\includegraphics[width=0.5cm]{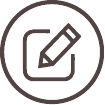}
\vskip-\intextsep
\end{wrapfigure}

\noindent
\textbf{\textsc{Modify}} revises a plan whose direction remains valuable but whose formulation needs adjustment.
To do so, the analyst first pauses the thread if it is still running, then clicks the modify button on the plan card and edits its description.
After confirmation, AdaLens updates the plan and keeps the thread available for relaunch under the revised formulation.

\vspace{1mm}
\begin{wrapfigure}[2]{l}{0.3cm}
\vskip-\intextsep
\includegraphics[width=0.5cm]{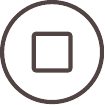}
\vskip-\intextsep
\end{wrapfigure}

\noindent
\textbf{\textsc{Terminate}} permanently stops a thread judged clearly unhelpful.
For pending or paused threads, termination takes effect immediately; for actively running threads, AdaLens stops the thread after its current operation completes.
The terminated thread is removed from subsequent execution.

\subsection{Implementation}
\label{sec:system-implementation}
We developed AdaLens as a web-based application using a client--server architecture, with a React-TS frontend and a Flask-based Python backend.
The backend exposes HTTP endpoints for execution and steering requests, and streams real-time events and generated artifacts to the frontend via Server-Sent Events (SSE).
The system is model-agnostic by design.
At the time of implementation, we used \texttt{gemini-3.1-pro-preview} for global orchestration and \texttt{gemini-3.1-flash-lite-preview} for workers through an OpenAI-compatible provider.
Further implementation details are provided in the supplementary material.

\section{Case Study}
This section presents two case studies and expert interviews to examine how AdaLens supports analysts in monitoring and steering agentic data analysis while pursuing their analytical tasks.
Together, the cases cover complementary forms of interactive oversight: observing process progression, findings, and data involvement through the storyline view, and interacting with its analytical elements for intention-level steering and execution-level control.
We invited two external analysts with domain expertise and practical experience using LLM-based tools for data analysis, neither of whom was involved in the design of AdaLens, to use the system to analyze datasets from their respective domains.
Each session lasted approximately one hour and followed a think-aloud protocol, after which we interviewed the experts to gather their feedback.
Although the sessions were bounded in duration, each analysis unfolded through successive stages as multiple analytical directions emerged and plans and findings accumulated.
This progression allowed us to examine how the experts monitored the evolving process, regulated execution, and steered subsequent analysis.


\begin{figure}[t]
  \centering
  \includegraphics[width=\columnwidth]{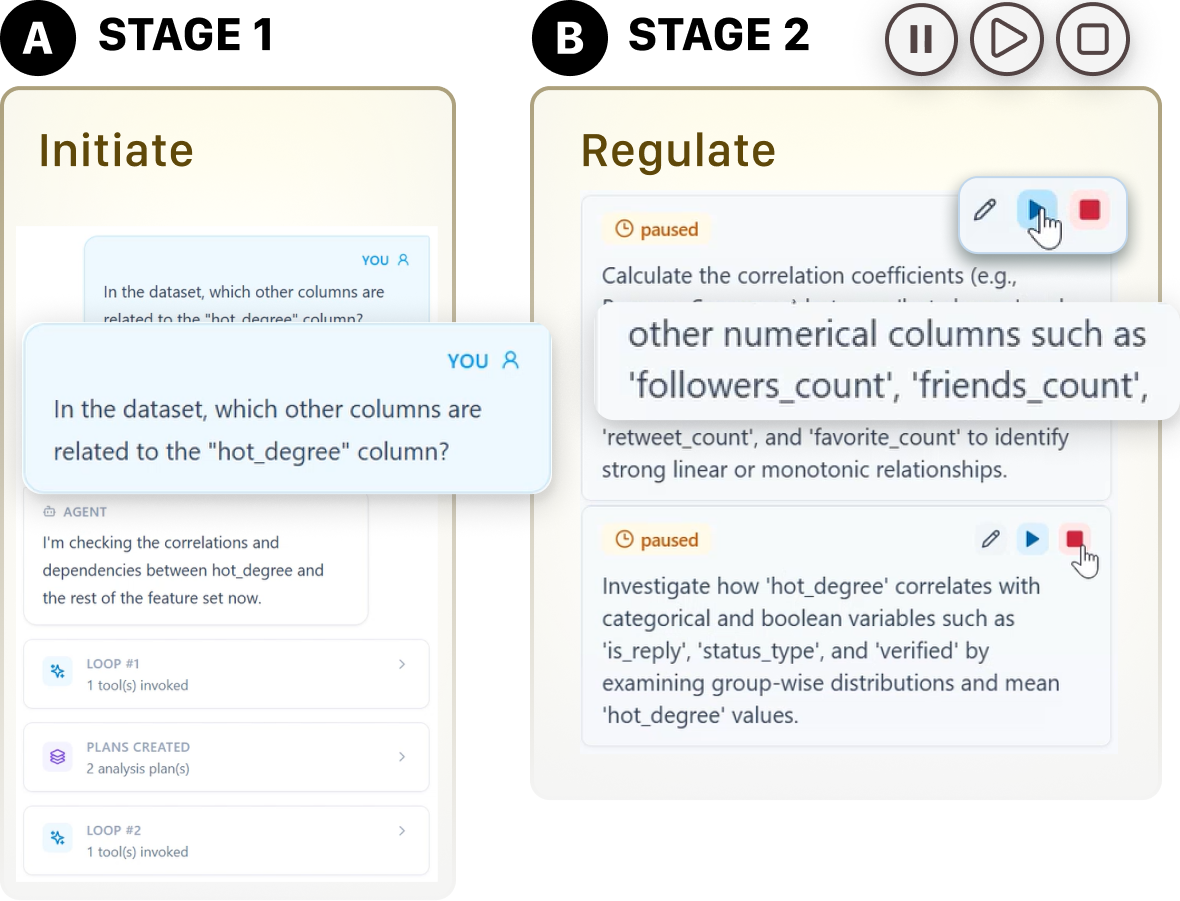}
  \caption{Case 1 (Stages 1--2): using AdaLens to initiate the investigation of the ambiguous \texttt{hot\_degree} metric in a Twitter dataset and regulate candidate plans.
  (A) ED starts from a focused chat request asking which columns relate to \texttt{hot\_degree}.
  (B) When two candidate plans appear in parallel with their execution states, ED pauses both, reviews them, terminates the categorical/boolean branch, and launches the branch on numerical variables.}
  \label{fig:case1-part1}
\end{figure}

\begin{figure*}[t]
  \centering
  \includegraphics[width=\textwidth]{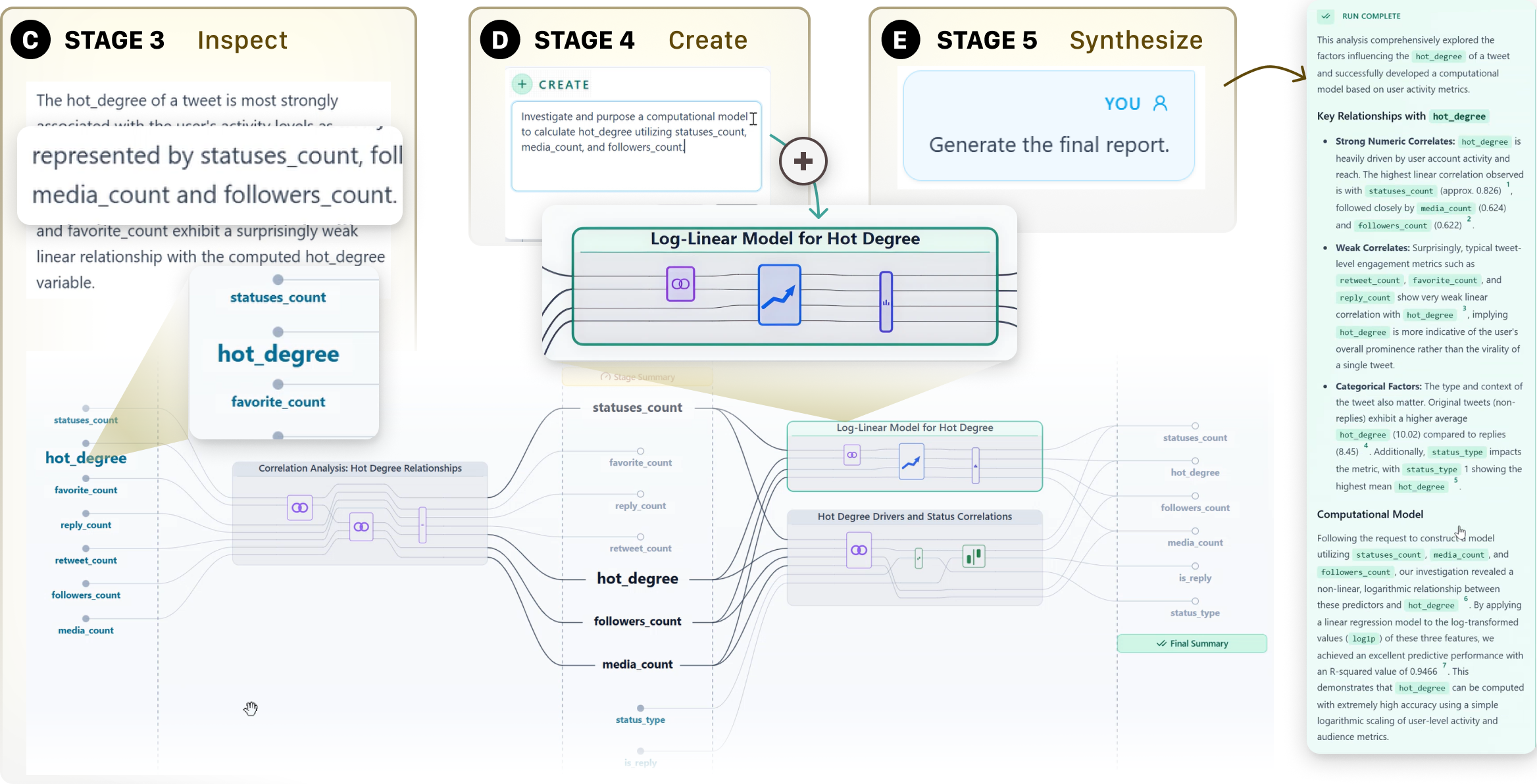}
  \caption{Case 1 (Stages 3--5): using AdaLens to inspect and extend the analysis of the ambiguous \texttt{hot\_degree} metric and synthesize the resulting findings.
  (C) Column label sizes convey the extent of each variable's involvement, highlighting \texttt{hot\_degree} as the analytical focus; the resulting summary identifies \texttt{statuses\_count}, \texttt{media\_count}, and \texttt{followers\_count} as its strongest related variables.
  (D) Based on these visible results, ED creates a targeted follow-up plan to model \texttt{hot\_degree} from these variables; the same four column characters continue into the follow-up step, which yields a log-linear model for \texttt{hot\_degree}.
  (E) AdaLens returns a final report that consolidates both the correlation findings and the resulting predictive model.}
  \label{fig:case1-part2}
\end{figure*}

\subsection{Case 1: Goal-Oriented Analysis of Social Media Data}
Expert D (male, ED) is a data analyst with two years of experience in social media data analysis.
For this case, ED provided a crawled Twitter dataset in which each row represents an individual tweet and includes the post text, posting context, user-account statistics, and engagement measures.
Among these variables was a metric labeled \texttt{hot\_degree}, whose meaning was ambiguous and whose relationship to other variables was unclear to him.
ED wanted to clarify what factors were associated with \texttt{hot\_degree} before relying on it further, as this metric could affect how the dataset was interpreted in subsequent analysis.
He therefore used AdaLens with a focused goal: to identify how \texttt{hot\_degree} related to the other columns in the dataset.

\textbf{Stage 1: Initiating analysis with a focused question.}
ED uploaded the dataset and typed his question in the chat view (\cref{fig:case1-part1}A): ``\textit{In the dataset, which other columns are related to the ``hot\_degree'' column?}''
The system first responded in the chat view, after which ED observed AdaLens begin creating plans for the analysis in the storyline view.

\textbf{Stage 2: Reviewing and regulating candidate plans.}
Once the system created and dispatched the initial plans, ED saw two new plan cards appear in parallel in the storyline view, together with their execution states (\cref{fig:case1-part1}B).
He first paused both threads to prevent the system from spending tokens on potentially unnecessary analysis while he reviewed them.
After inspecting the two candidate plans, he terminated the branch on categorical and boolean variables, because he wanted to first focus on the numerical variables that were more likely to directly explain \texttt{hot\_degree}.
ED then launched the remaining plan and monitored its progress as the card state transitioned from analyzing to summarizing, eventually producing a summary card.

\textbf{Stage 3: Inspecting data involvement and the summary.}
After the storyline updated (\cref{fig:case1-part2}C), ED noticed that the label for \texttt{hot\_degree} was displayed at the largest size among the columns in this step, indicating that the analysis remained centered on his target variable.
He then inspected the resulting summary, which reported that \texttt{hot\_degree} was most strongly related to \texttt{statuses\_count}, followed by \texttt{media\_count} and \texttt{followers\_count}.
The labels for these three columns were also relatively large, making their prominent involvement in this step readily visible.

\textbf{Stage 4: Creating a targeted follow-up plan.}
Based on this finding and the three identified variables, ED wanted to move beyond correlation and understand how \texttt{hot\_degree} might be computationally derived.
Because the main analytical run was still ongoing, he chose not to redirect it by sending a new prompt through the chat view.
Instead, he right-clicked in the storyline view to trigger the \textsc{Create} interaction and entered a more targeted request (\cref{fig:case1-part2}D): ``\textit{Investigate and propose a computational model to calculate hot\_degree utilizing statuses\_count, media\_count, and followers\_count.}''
After confirmation, AdaLens created a new plan card and dispatched it within the current execution batch.
When this plan completed, the resulting summary card reported:
``\textit{A linear regression model using log-transformed values of statuses\_count, media\_count, and followers\_count effectively predicts hot\_degree with an R-squared of 0.9466.}''
ED observed that the column characters for these four variables continued into the follow-up step, showing that the targeted analysis remained grounded in the same data dimensions as the preceding correlation analysis.

\textbf{Stage 5: Requesting a final synthesis.}
At this point, ED had identified the variables most strongly related to \texttt{hot\_degree} and obtained a plausible computational model for estimating it.
He returned to the chat view and issued a final summarization request, asking AdaLens to synthesize the overall findings.
In response, the system generated a global report (\cref{fig:case1-part2}E) that consolidated the earlier results into a coherent account of how \texttt{hot\_degree} related to the other columns in the dataset.

\begin{figure*}[t]
  \centering
  \includegraphics[width=\textwidth]{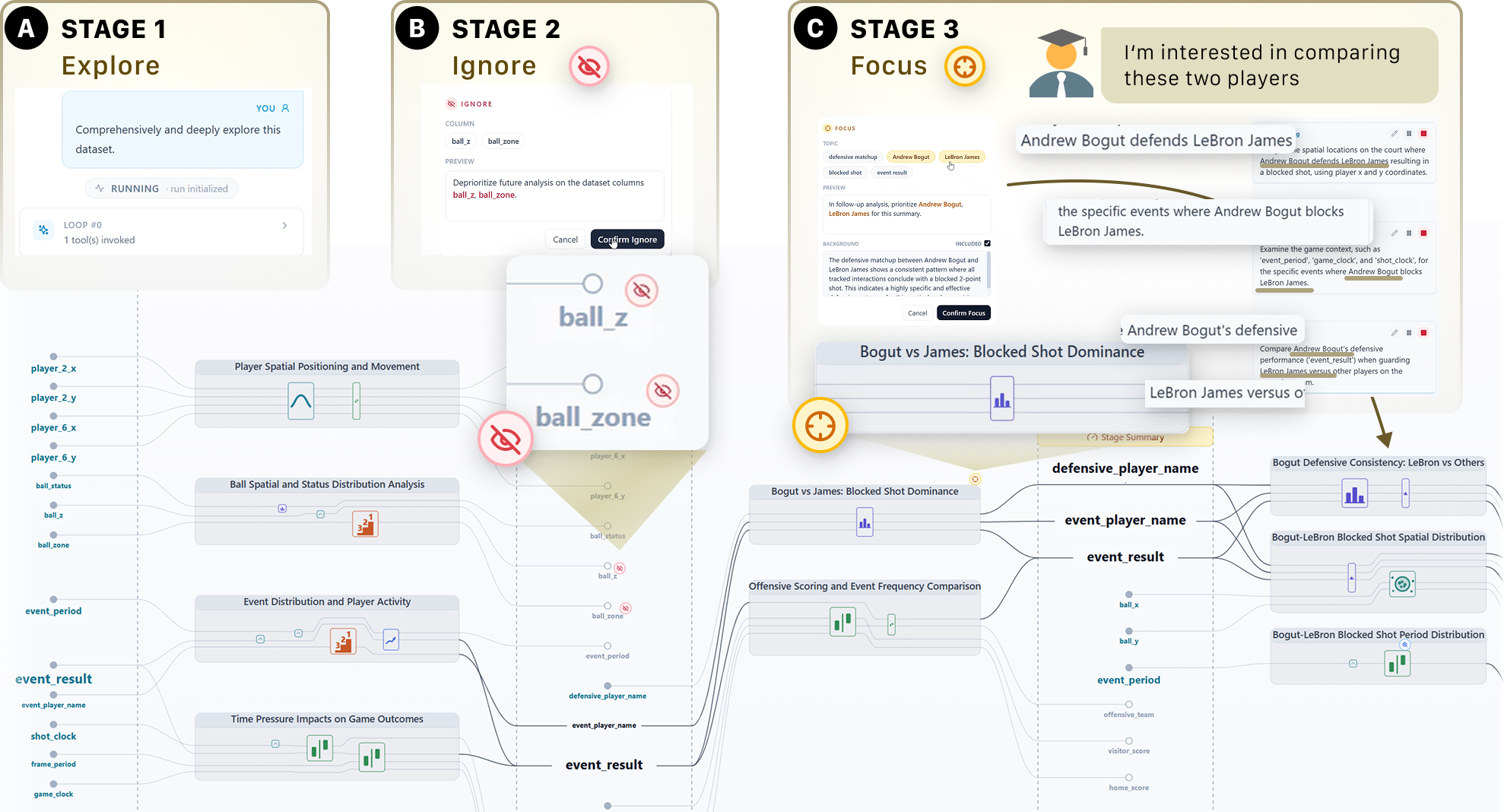}
  \caption{Case 2 (Stages 1--3): using AdaLens to explore an NBA spatiotemporal tracking dataset in an open-ended manner and steer the analysis toward a promising matchup.
  (A) EE begins with a broad exploration request, after which the storyline presents early summaries, atomic insights, and the columns involved in them.
  (B) EE notices two frame-level ball-coordinate columns that are not of particular interest and applies \textsc{Ignore} to \texttt{ball\_zone} and \texttt{ball\_z} to deprioritize them.
  (C) He then applies \textsc{Focus} to a promising summary card about Bogut and James, steering subsequent plans toward this matchup.}
  \label{fig:case2-part1}
\end{figure*}

\begin{figure*}[t]
  \centering
  \includegraphics[width=\textwidth]{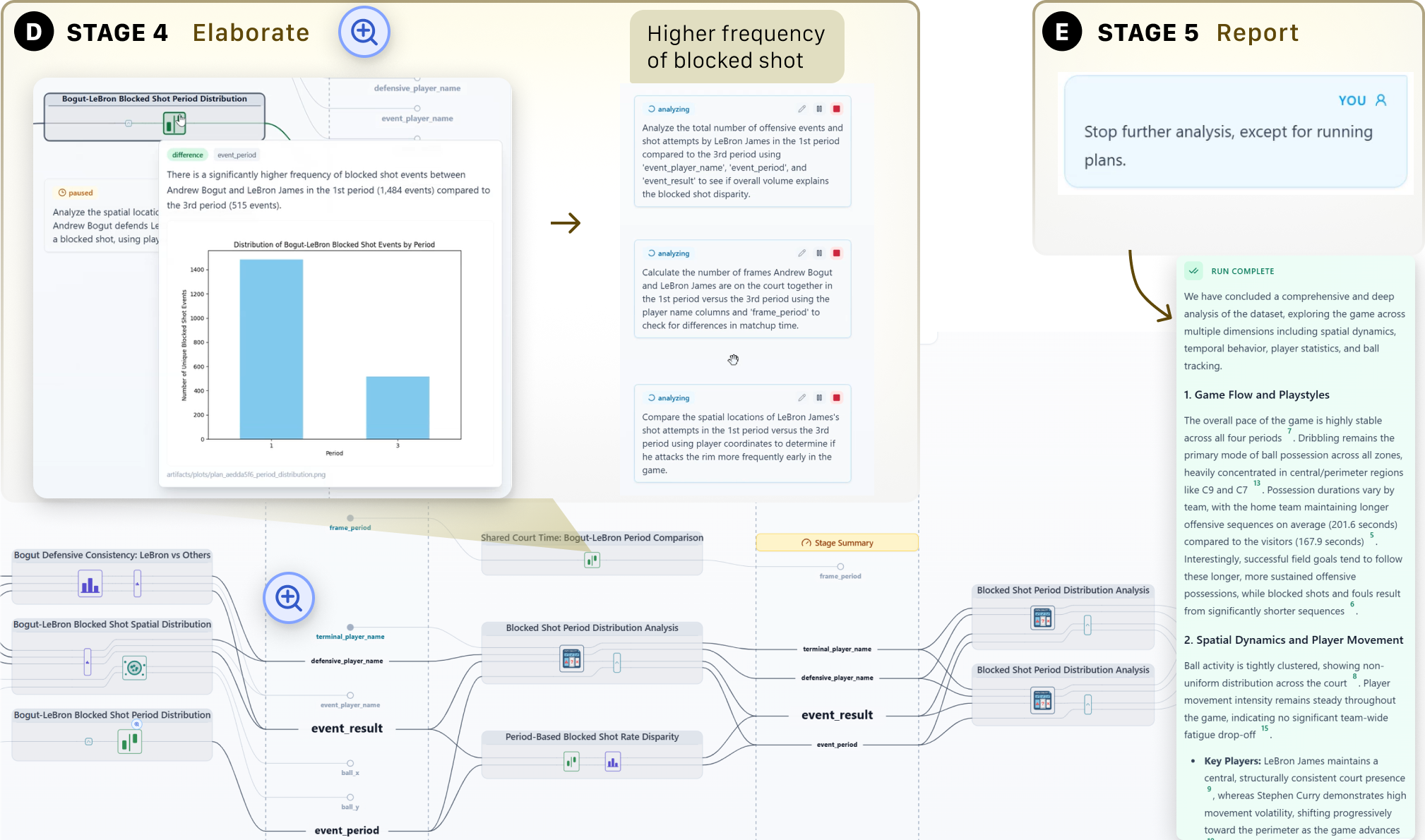}
  \caption{Case 2 (Stages 4--5): using AdaLens to elaborate a key insight and generate a final report.
  (D) Within the resulting summaries, an atomic insight reveals that blocked-shot interactions are more frequent in the first period than in the third, prompting EE to apply \textsc{Elaborate} to it for a deeper explanation of this temporal pattern.
  (E) AdaLens finally generates a report that consolidates the exploratory findings into a coherent interpretation of the game.}
  \label{fig:case2-part2}
\end{figure*}

\subsection{Case 2: Exploratory Analysis of NBA Game Data}
Expert E (male, EE) is a sports data analyst with three years of experience in basketball game data analysis.
For this case, EE provided an NBA spatiotemporal tracking dataset from the 2015--2016 Christmas Day game between the Cavaliers and the Warriors, organized by possession while retaining fine-grained spatial records.
Because the dataset is high-dimensional and tactically rich, EE did not begin with a single narrowly defined question.
Instead, he used AdaLens for exploratory analysis to surface potentially meaningful patterns in the game.

\textbf{Stage 1: Starting with open-ended exploration.}
EE uploaded the dataset and initiated an open-ended analysis: ``\textit{Comprehensively and deeply explore this dataset.}'' 
As the initial plans completed, the storyline view (\cref{fig:case2-part1}A) began to accumulate summary cards and atomic insights connected to data columns.
This allowed EE to survey the emerging findings, potential analytical directions, and involved columns together during the early exploratory phase.

\textbf{Stage 2: Ignoring low-interest ball-coordinate columns.}
Among the column characters associated with the emerging findings, EE noticed several that represented frame-level ball coordinates, which were not of particular interest to his exploration.
He therefore activated the \textsc{Ignore} tool and selected two such column characters, labeled \texttt{ball\_zone} and \texttt{ball\_z}, directly in the storyline view (\cref{fig:case2-part1}B).
After confirmation, the two columns were marked with the \textsc{Ignore} visual indicator in the storyline view, and AdaLens automatically sent a steering prompt in the chat view: ``\textit{Deprioritize future analysis of dataset columns ball\_zone and ball\_z.}''
In subsequent rounds, these two columns no longer appeared in the storyline view, confirming that the system had successfully deprioritized them.

\textbf{Stage 3: Focusing on a promising summary.}
Among the accumulated findings, EE noticed a summary card labeled ``\textit{Bogut vs. James: Blocked-shot dominance}.''
This immediately drew his attention to the matchup between Bogut and James, and he decided to pursue this lead to better understand their head-to-head performance.
EE applied the \textsc{Focus} tool to this summary (\cref{fig:case2-part1}C).
A temporary card appeared, prompting him to select topics from the summary.
After choosing ``\textit{Bogut}'' and ``\textit{James}'', EE saw an automatically generated preview prompt, ``\textit{In follow-up analysis, prioritize Bogut and James for this summary}'', with the selected summary included as contextual background by default.
After he confirmed, AdaLens sent the prompt to the chat view, and in the next round of analysis, all proposed plans centered on Bogut and James.

\textbf{Stage 4: Elaborating a key insight to understand why it emerged.}
Within the resulting Bogut--James summaries, EE inspected an atomic insight stating, ``\textit{Blocked-shot events between Andrew Bogut and LeBron James occur substantially more often in the first period than in the third period}.''
Rather than stopping at this observation, EE wanted to understand why this temporal pattern emerged.
He applied the \textsc{Elaborate} tool to the insight glyph (\cref{fig:case2-part2}D) to request a deeper explanation of this specific finding.
After confirmation, AdaLens sent the prompt to the chat view, and the resulting follow-up analysis provided a more detailed account of the pattern.
Through this progression, EE acted successively on visible columns, a summary card, and an atomic insight to deprioritize low-interest variables, pursue a promising matchup, and deepen a specific finding.
These interactions moved the analysis from identifying the salient Bogut--James matchup to developing a more detailed explanation of why their blocked-shot interactions were concentrated early in the game.

\textbf{Stage 5: Consolidating the exploratory findings.}
After the exploration, EE returned to the chat view and requested a final synthesis of the key patterns surfaced during the run.
AdaLens generated a global summary report (\cref{fig:case2-part2}E) that consolidated the findings into a coherent interpretation of the game.

\subsection{Expert Interviews}
After completing their analyses, we conducted one-on-one semi-structured interviews with ED and EE to gather their perspectives on AdaLens. 
We summarize their feedback in three aspects below.

\textbf{Visual Design and Interactions.}
Both experts highlighted the storyline view as the most valuable component of AdaLens.
EE felt it provided a clear overview of process progression and step-level findings, while ED particularly valued the persistent column traces and size encoding, explaining that ``\textit{this size encoding is important}'' because it helped him see ``\textit{what is mainly being analyzed here.}''
Both also valued the direct steering interactions.
EE felt that \textsc{Ignore}, \textsc{Focus}, and \textsc{Elaborate} made it easy to continue from visible results without composing new instructions from scratch, and ED called this form of direct manipulation ``\textit{definitely better}'' than pure text control.


\textbf{Usability in real-world data analysis.}
Both experts familiarized themselves with the interface quickly after a brief tutorial and did not request additional help on system usage during the case studies, suggesting a low learning barrier.
ED appreciated the fine-grained control over plan threads, such as terminating one thread without stopping the entire run, noting that AdaLens offered stronger support for maintaining context than his current workflow of tracking agent behavior through evolving text outputs.
EE felt the system suited exploratory work well and indicated he would use such a system in practice, though he noted that analysts who rely heavily on code inspection may want stronger support for following the code-level analytical process.


\textbf{Suggestions for Improvement.}
ED noted that related results can become difficult to connect when they are far apart along the timeline, explaining that ``\textit{it becomes easy to get lost in the middle}'' and suggesting more explicit visual links between connected results.
EE raised a similar concern, suggesting that making inter-branch relationships more explicit, such as whether results are correlated or causally linked, would support better context management across longer runs.
We discuss the broader challenge of maintaining readability and orientation in longer runs as a limitation in \cref{sec:discussion}.


\section{User Study}
We conducted a task-based user study to evaluate the usability of AdaLens and collect feedback on participants' experiences with the system.

\subsection{Experiment Settings}

We recruited 12 undergraduate and graduate students (P1--P12; 8 male, 4 female; aged 20--26) with prior LLM and data analysis experience through a university forum.
To ensure diverse perspectives, participants were drawn from varied academic disciplines across STEM (e.g., computer science, engineering), social sciences, and humanities.
Each participant was compensated US\$12 for their time.


\textbf{Procedure and Tasks.}
Our study procedure consisted of four stages:
\uline{\textit{(1) Tutorial and Warm-up (15 mins)}}.
We first introduced the study background and AdaLens, then guided participants through a warm-up task with a video game sales dataset~\cite{smith2022videogamesales} to familiarize them with the interface and key functions.
\uline{\textit{(2) Tasks (15 mins)}}.
Participants were asked to complete sequential tasks using AdaLens to analyze a student performance dataset~\cite{student_performance_320}.
Although limited to 15 minutes, the task phase was designed to cover the core monitoring and steering activities that arise in an evolving, multi-step run.
Over the course of the run, participants needed to inspect the evolving analytical process, regulate concurrent plans, steer subsequent analysis based on visible results, and review the final output.
The detailed task descriptions are provided in the supplementary material.
\uline{\textit{(3) Free-form exploration (10 mins)}}. After completing the tasks, participants were encouraged to freely use and explore the system.
\uline{\textit{(4) Semi-structured interview (10 mins)}}. 
Participants completed a SUS questionnaire~\cite{JBrooke1996sus} followed by a semi-structured interview regarding their experience.
All sessions were screen-recorded and transcribed.

\subsection{Findings}
The average SUS score was 87.08, as shown in~\cref{fig:sus}, exceeding the $A^+$ rating threshold of 84.1~\cite{measuringu_sus}.
Following the SUS factor-structure study~\cite{lewis2009factorsus}, we also separately calculated the system's usability and learnability scores, which were 87.76 and 84.38, respectively.
These scores suggest that participants found AdaLens highly usable and easy to learn.
Furthermore, we analyzed the interview transcripts through thematic coding and reviewed the recordings for complementary observations of participants' task performance and behavior.
The main findings are as follows:

\begin{figure}[t]
  \centering
  \includegraphics[width=\columnwidth]{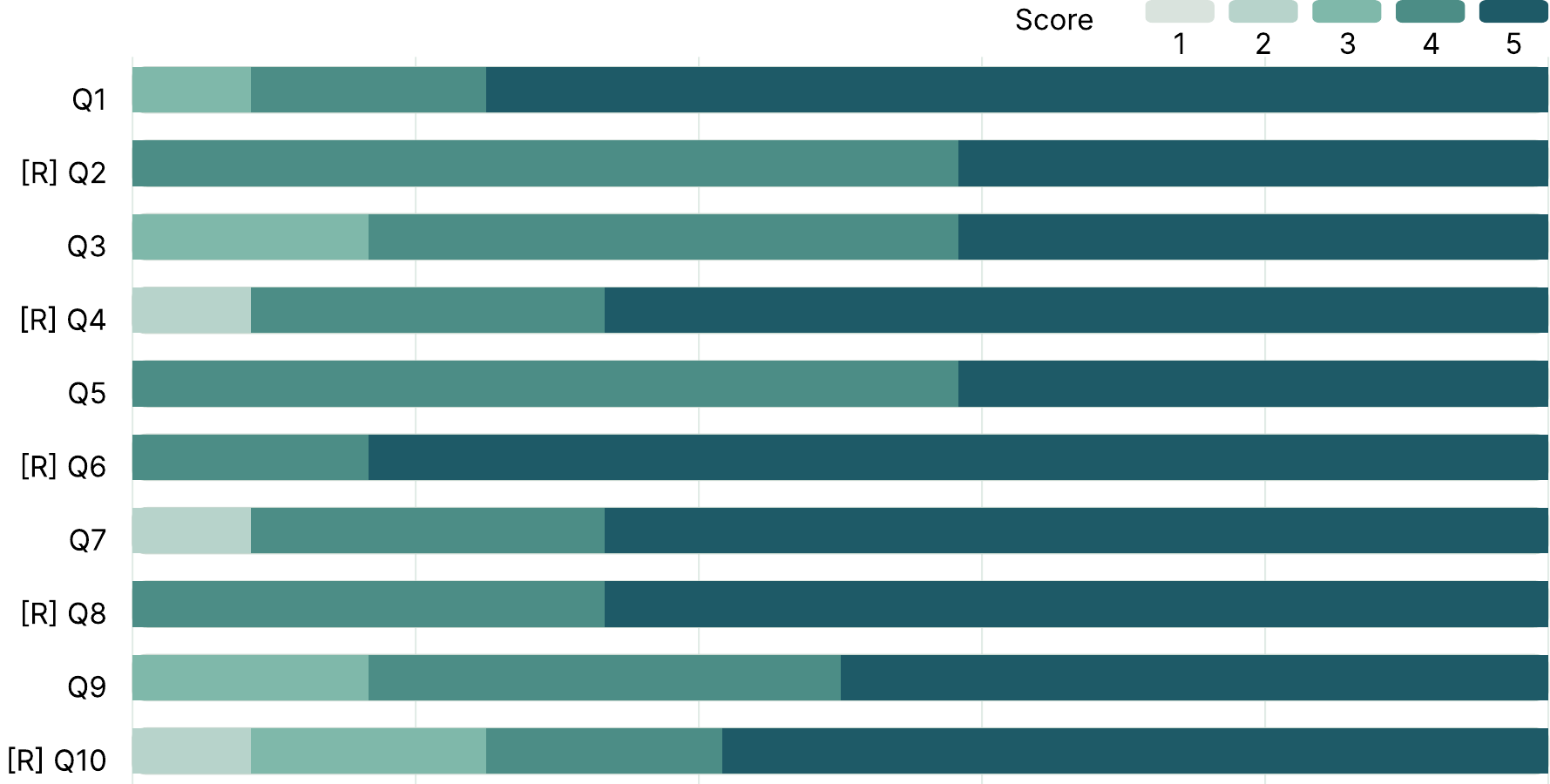}
  \caption{Usability results from a 5-point SUS questionnaire~\cite{JBrooke1996sus}.
  Negative questions (Q2, Q4, Q6, Q8, and Q10) are shown inverted for consistency.}
  \label{fig:sus}
\end{figure}

\textbf{System usability and learnability.}
All participants (12/12) gave overall positive feedback on usability.
They commonly described it as ``\textit{useful}'', ``\textit{intuitive}'', ``\textit{clear}'', and ``\textit{easy to use}''.
Several participants (P2, P4, P10, P11, P12) particularly appreciated that AdaLens was feature-rich without feeling overly complex.
As P2 said, ``\textit{This system has quite a few features, but they are well integrated and clearly defined, so I do not find it very complex.}''
Regarding learnability, most participants (10/12) explicitly stated that AdaLens was easy to learn.
As P5 noted, ``\textit{I think this (AdaLens) was very quick to get started with, and it was easy to understand what each of its functions meant.}''
Nevertheless, several participants (P1, P8, P10, P12) pointed out that diverse insight glyphs and steering interactions could take some time to fully master.
Observed task performance provided complementary context: 9/12 participants completed the prescribed task sequence using only the provided task instructions, whereas the remaining three required additional guidance for some operations.

\textbf{Observability of the analytical process.}
All participants (12/12) commented positively on AdaLens's support for observing the analytical process.
Many participants (P2--P4, P7, P10, P11, P12) especially appreciated how the storyline view made the stepwise process and intermediate findings visible.
As P7 noted, ``\textit{I could clearly see how it analyzed the data step by step, how many steps there were, and what insights it produced along the way.}''
Many participants (8/12) also highlighted the value of being able to inspect supporting evidence and trace the reasoning process, rather than relying solely on final textual answers.
For example, P4 explained, ``\textit{I could clearly see how it reasoned through each step instead of just throwing out a result, and I could immediately tell whether its direction had gone off track.}''
Several participants (P3, P4, P6, P11) further noted that this visibility increased their trust, confidence, or sense of control over the analytical process.
Participants (P2, P5, P7, P9, P10) also appreciated how the storyline view made data-column involvement visible, helping them track which variables were used, how they were recombined across steps, and where a particular finding came from.
The recordings also captured participants actively using these inspection capabilities during free-form exploration: 11/12 participants independently revisited previously generated atomic insights for closer inspection and then examined their associated data columns or supporting evidence.

\textbf{Steerability of the analytical process.}
Many participants (9/12) expressed positive sentiments regarding AdaLens's direct steering interactions.
They particularly appreciated being able to redirect the ongoing analysis by interacting with visible analytical elements.
Several participants (P1, P2, P4, P6, P8, P9) described these interactions as useful for refining analysis goals on the fly.
P6 summarized this benefit by noting, ``\textit{I could adjust the analysis directly in the graphical interface, such as focusing on a specific insight, which was much more convenient than writing prompts myself.}''
Participants (P2, P8, P9) also noted that the steering actions made the relationship between intervention and subsequent analysis more visible, helping them understand how their guidance affected later steps.
The recordings also captured participants actively using these steering interactions during free-form exploration: 9/12 participants independently identified analytical directions of interest in the storyline and used \textsc{Focus} or \textsc{Elaborate} to initiate follow-up analysis.
Among them, five further edited the generated steering prompts to add more specific instructions.
However, some participants (P7, P8) pointed out that using these interactions effectively still required users to keep up with the agent's progress and maintain a clear understanding of the current analytical structure.

\textbf{Perceived contrast with chat-based workflows.}
In this self-reported comparison, all participants (12/12), drawing on their prior LLM-use experience, felt that AdaLens would better support long-running agentic data analysis workflows than a chat-only interface.
Participants (P1, P2, P4, P6, P7, P8, P9) emphasized that it supported a different mode of work from chat-only systems.
P8 described this difference succinctly: ``\textit{The interaction shifts from a cycle of asking, getting a result, and then adding more conditions, to adjusting the analysis in real time through the graphical interface.}''
Several participants (P1, P2, P4, P7, P8, P9) noted that this reduced the need to reconstruct context or rewrite extensive prompts when continuing the analysis.
Participants (P3, P4, P11, P12) further remarked that, relative to their prior experience with pure chat interfaces, AdaLens felt more grounded and trustworthy because they could inspect the running analysis rather than diagnose problems only after a final answer was produced.

\textbf{Suggestions.}
Participants also proposed several concrete suggestions for AdaLens, mainly around more explicit hints for visual elements and interactions (P1, P4, P8, P10, P11).
Others focused on scalability and layout organization, recommending more compact layouts to accommodate numerous analytical elements (P2, P9), reduced visual clutter (P2, P5, P7, P9), and clearer ordering of important items (P10, P12).
These suggestions highlight opportunities to further improve the learnability and usability of the system.

\section{Discussion}
\label{sec:discussion}
In this section, we reflect on the implications and lessons learned from AdaLens, then discuss its limitations and directions for future work.

\textbf{Implications.}
We discuss this from three perspectives:

\textit{Interaction paradigm.}
Our evaluation suggests that long-running agentic data analysis would benefit from interfaces beyond prompt--response exchanges: analysts tracked evolving run state, examined accumulated context through the storyline, and intervened mid-run, indicating that continuous monitoring and in-situ redirection are important interaction requirements for this setting.


\textit{Techniques.}
This work extends classical storyline visualization by framing long-running agentic data analysis as an evolving narrative: plan threads are represented by plan cards that become summary cards upon completion, marking analytical events; atomic insights represent fine-grained findings nested within completed summaries; and data columns act as persistent characters whose trajectories connect the analytical events and findings that involve them.
This mapping unifies analytical elements of different granularity in a temporal representation that preserves data-grounded analytical lineage across steps.
To support ongoing runs rather than post-hoc histories, the progressive layout incrementally extends the storyline as new analytical elements arrive, preserving previously rendered context while maintaining visual clarity and compactness.
Across the case studies, the storyline served not only as a representation for monitoring evolving analytical elements and their data-grounded lineage, but also as an interaction surface for steering the ongoing analysis.
Grounding monitoring and steering in the same visible elements allowed analysts to translate what they noticed into contextualized actions.


\textit{Applicability.}
Both the focused social-media case and the exploratory NBA case suggest that combining conversational framing with structured visual oversight can support both directed follow-up and open-ended exploration.
This combination may generalize to other long-running agentic workflows, such as automated report generation or iterative model-building pipelines, where analysts similarly need to monitor evolving state and intervene selectively.


\textbf{Lessons learned.}
Four lessons emerged from the design of AdaLens.
First, effective oversight benefits from semantically meaningful, multi-granular abstractions rather than exhaustive exposure of low-level traces, because analysts naturally reasoned in terms of plans, findings, and data columns.
Second, observability and steerability become more coherent when they are built on the same analytical objects, allowing users to move directly from understanding to intervention without restating accumulated context elsewhere.
Third, because long-running analyses expand incrementally and branch over time, layout stability is not merely presentational but essential for preserving orientation and maintaining the visibility of analytical lineage as new findings arrive.
Finally, richer oversight must balance expressiveness with learnability and scalability, as participants' feedback on insight glyphs, steering operations, and storyline density showed that expressive interfaces also require clear guidance, visual simplification, and scalable organization.

\textbf{Limitations and future work.}

(1) Our current evaluation offers an initial assessment of AdaLens through case studies and a task-based user study.
While the results indicate promising support for monitoring and steering evolving, multi-step agentic data analyses, longitudinal, in-situ, and comparative studies involving analytical runs of longer duration are still needed to examine how these benefits carry over to extended real-world use.

(2) AdaLens is currently instantiated in a structured-data setting, and its storyline deliberately uses recurring data columns as a persistent connective layer for representing analytical lineage.
Shared columns reveal common data involvement, but they are limited in their ability to explicitly reveal other relationships among analytical elements, such as semantic relatedness, causal dependence, shared conclusions, or data provenance.
Future work can extend this foundation with complementary relationships based on hypotheses, assumptions, operations, models, results, or finer-grained row-level provenance, and examine how these abstractions generalize to richer data modalities and workflows.

(3) As runs become longer and denser, maintaining readability and orientation remains an important design challenge.
The user feedback on more compact layouts and clearer guidance suggests opportunities to further improve scalability and usability, while the quality of the interface also depends on the quality of backend summarization.
Future work could explore multiscale storyline abstractions, richer interaction feedback, and tighter integration between interface design and backend support.


\section{Conclusion}
In this paper, we presented AdaLens, an interactive system for monitoring and steering long-running agentic data analysis.
Through close collaboration with three researchers experienced in agentic data analysis, we formulated this problem and identified two key design challenges.
We designed AdaLens around a storyline-based representation with a progressive layout, together with steering interactions grounded in analytical elements.
Through two case studies and a user study, we examined how AdaLens supports analysts in monitoring and steering long-running agentic data analysis.




\bibliographystyle{IEEEtran}
\bibliography{references-v0.1}

@String{LACC    = {(Last accessed: Mar 27, 2026)}}

@String{UIST    = {Proc. ACM Symp. User Interface Softw. Technol.}}

@String{CHI     = {Proc. ACM CHI Conf. Hum. Factors Comput. Syst.}}

@String{IUI     = {Proc. ACM Int. Conf. Intell. User Interfaces}}

@String{TVCG    = {IEEE Trans. Vis. Comput. Graph.}}

@String{ACL     = {Proc. Annu. Meet. Assoc. Comput. Linguist.}}

@String{NAACL   = {Proc. Annu. Conf. North Am. Chapter Assoc. Comput. Linguist.}}

@String{EMNLP   = {Proc. Conf. Empir. Methods Nat. Lang. Process.}}

@String{FINDACL = {Find. Assoc. Comput. Linguist.}}

@String{ICLR    = {Proc. Int. Conf. Learn. Represent.}}

@String{ICML    = {Proc. Int. Conf. Mach. Learn.}}

@String{CANDC   = {Proc. Conf. Creat. Cogn.}}

@String{UISTADJ = {Adjun. Proc. ACM Symp. User Interface Softw. Technol.}}

@String{CGANDA  = {IEEE Comput. Graph. Appl.}}

@String{HCD     = {Proc. Int. Conf. Hum. Centered Des.}}

@String{CGF     = {Comput. Graph. Forum}}

@article{zhu2025dataagents,
  author  = {Yizhang Zhu and Liangwei Wang and Chenyu Yang and Xiaotian Lin and Boyan Li and Wei Zhou and Xinyu Liu and Zhangyang Peng and Tianqi Luo and Yu Li and Chengliang Chai and Chong Chen and Shimin Di and Ju Fan and Ji Sun and Nan Tang and Fugee Tsung and Jiannan Wang and Chenglin Wu and Yanwei Xu and Shaolei Zhang and Yong Zhang and Xuanhe Zhou and Guoliang Li and Yuyu Luo},
  title   = {A Survey of Data Agents: Emerging Paradigm or Overstated Hype?},
  journal = {arXiv preprint arXiv:2510.23587},
  year    = {2025},
  doi     = {10.48550/arXiv.2510.23587}
}

@misc{vizgpt2024,
  key          = {VizGPT},
  title        = {{AI Data Visualization \& Analytics Platform}},
  howpublished = {{vizGPT}},
  note         = {Available: \url{https://vizgpt.ai/} } # LACC
}

@misc{openai2022chatgpt,
  author       = {{OpenAI}},
  title        = {Introducing {ChatGPT}},
  howpublished = {{OpenAI}},
  month        = nov,
  year         = {2022},
  note         = {Available: \url{https://openai.com/index/chatgpt/} } # LACC
}

@misc{juliusai,
  author       = {{Julius AI}},
  title        = {{Julius AI}: Excel, Slides, Tasks with AI},
  howpublished = {{Julius AI}},
  note         = {Available: \url{https://julius.ai/} } # LACC
}

@inproceedings{dibia2023lida,
  author    = {Dibia, Victor},
  title     = {{LIDA}: A Tool for Automatic Generation of Grammar-Agnostic Visualizations and Infographics using Large Language Models},
  booktitle = ACL,
  year      = {2023},
  pages     = {113--126},
  doi       = {10.18653/v1/2023.acl-demo.11}
}

@article{chen2025prompt4vis,
  author  = {Li, Shuaimin and Chen, Xuanang and Song, Yuanfeng and Song, Yunze and Zhang, Chen},
  title   = {{Prompt4Vis}: Prompting Large Language Models with Example Mining and Schema Filtering for Tabular Data Visualization},
  journal = {arXiv preprint arXiv:2402.07909},
  year    = {2024},
  doi     = {10.48550/arXiv.2402.07909}
}

@article{wang2025noteex,
  author  = {Payandeh, Mohammad Hasan and Yuan, Lin-Ping and Zhao, Jian},
  title   = {{NoteEx}: Interactive Visual Context Manipulation for {LLM}-Assisted Exploratory Data Analysis in Computational Notebooks},
  journal = {arXiv preprint arXiv:2511.07223},
  year    = {2025},
  doi     = {10.48550/arXiv.2511.07223}
}

@article{wang2025dagent,
  author  = {Xu, Wenyi and Mao, Yuren and Zhang, Xiaolu and Zhang, Chao and Dong, Xuemei and Zhang, Mengfei and Gao, Yunjun},
  title   = {{DAgent}: A Relational Database-Driven Data Analysis Report Generation Agent},
  journal = {arXiv preprint arXiv:2503.13269},
  year    = {2025},
  doi     = {10.48550/arXiv.2503.13269}
}

@article{agenticdata2025,
  author  = {Sun, Ji and Li, Guoliang and Zhou, Peiyao and Ma, Yihui and Xu, Jingzhe and Li, Yuan},
  title   = {{AgenticData}: An Agentic Data Analytics System for Heterogeneous Data},
  journal = {arXiv preprint arXiv:2508.05002},
  year    = {2025},
  doi     = {10.48550/arXiv.2508.05002}
}

@inproceedings{gu2024analysts,
  author    = {Gu, Ken and Shang, Ruoxi and Althoff, Tim and Wang, Chenglong and Drucker, Steven M.},
  title     = {How Do Analysts Understand and Verify {AI}-Assisted Data Analyses?},
  booktitle = CHI,
  year      = {2024},
  pages     = {1--22},
  doi       = {10.1145/3613904.3642497}
}

@inproceedings{kim2024analysts,
  author    = {Gu, Ken and Grunde-McLaughlin, Madeleine and McNutt, Andrew and Heer, Jeffrey and Althoff, Tim},
  title     = {How Do Data Analysts Respond to {AI} Assistance? {A Wizard-of-Oz} Study},
  booktitle = CHI,
  year      = {2024},
  pages     = {1--22},
  doi       = {10.1145/3613904.3641891}
}

@inproceedings{subramonyam2024bridging,
  author    = {Subramonyam, Hari and Pea, Roy and Pondoc, Christopher and Agrawala, Maneesh and Seifert, Colleen},
  title     = {Bridging the Gulf of Envisioning: Cognitive Challenges in Prompt Based Interactions with {LLMs}},
  booktitle = CHI,
  pages     = {1--19},
  year      = {2024},
  doi       = {10.1145/3613904.3642754}
}

@inproceedings{tankelevitch2024metacognitive,
  author    = {Tankelevitch, Lev and Kewenig, Viktor and Simkute, Auste and Scott, Ava Elizabeth and Sarkar, Advait and Sellen, Abigail and Rintel, Sean},
  title     = {The Metacognitive Demands and Opportunities of Generative {AI}},
  booktitle = CHI,
  pages     = {1--24},
  year      = {2024},
  doi       = {10.1145/3613904.3642902}
}

@article{shen2024bidirectional,
  author  = {Hua Shen and Tiffany Knearem and Reshmi Ghosh and Kenan Alkiek and Kundan Krishna and Yachuan Liu and Ziqiao Ma and Savvas Petridis and Yi-Hao Peng and Li Qiwei and Sushrita Rakshit and Chenglei Si and Yutong Xie and Jeffrey P. Bigham and Frank Bentley and Joyce Chai and Zachary Lipton and Qiaozhu Mei and Rada Mihalcea and Michael Terry and Diyi Yang and Meredith Ringel Morris and Paul Resnick and David Jurgens},
  title   = {Position: Towards Bidirectional Human-{AI} Alignment},
  journal = {arXiv preprint arXiv:2406.09264},
  year    = {2024},
  doi     = {10.48550/arXiv.2406.09264}
}

@article{feng2024xnli,
  author  = {Feng, Yingchaojie and Wang, Xingbo and Pan, Bo and Wong, Kam Kwai and Ren, Yi and Liu, Shi and Yan, Zihan and Ma, Yuxin and Qu, Huamin and Chen, Wei},
  title   = {{XNLI}: Explaining and Diagnosing {NLI}-Based Visual Data Analysis},
  journal = TVCG,
  volume  = {30},
  number  = {7},
  pages   = {3813--3827},
  year    = {2024},
  doi     = {10.1109/TVCG.2023.3240003}
}

@inproceedings{wu2022aichains,
  author    = {Wu, Tongshuang and Terry, Michael and Cai, Carrie Jun},
  title     = {{AI} Chains: Transparent and Controllable Human-{AI} Interaction by Chaining Large Language Model Prompts},
  booktitle = CHI,
  pages     = {1--22},
  year      = {2022},
  doi       = {10.1145/3491102.3517582}
}

@inproceedings{cai2024lowcodellm,
  author    = {Cai, Yuzhe and Mao, Shaoguang and Wu, Wenshan and Wang, Zehua and Liang, Yaobo and Ge, Tao and Wu, Chenfei and You, Wang and Song, Ting and Xia, Yan and Duan, Nan and Wei, Furu},
  title     = {Low-code {LLM}: Graphical User Interface over Large Language Models},
  booktitle = NAACL,
  pages     = {12--25},
  year      = {2024},
  doi       = {10.18653/v1/2024.naacl-demo.2}
}

@inproceedings{vaithilingam2024dynavis,
  author    = {Vaithilingam, Priyan and Glassman, Elena L. and Inala, Jeevana Priya and Wang, Chenglong},
  title     = {{DynaVis}: Dynamically Synthesized {UI} Widgets for Visualization Editing},
  booktitle = CHI,
  pages     = {1--17},
  year      = {2024},
  doi       = {10.1145/3613904.3642639}
}

@inproceedings{masson2024directgpt,
  author    = {Masson, Damien and Malacria, Sylvain and Casiez, G{\'e}ry and Vogel, Daniel},
  title     = {{DirectGPT}: A Direct Manipulation Interface to Interact with Large Language Models},
  booktitle = CHI,
  pages     = {1--16},
  year      = {2024},
  doi       = {10.1145/3613904.3642462}
}

@article{chen2024sketchthengenerate,
  author  = {Zhu-Tian, Chen and Xiong, Zeyu and Yao, Xiaoshuo and Glassman, Elena},
  title   = {Sketch Then Generate: Providing Incremental User Feedback and Guiding {LLM} Code Generation through Language-Oriented Code Sketches},
  journal = {arXiv preprint arXiv:2405.03998},
  year    = {2024},
  doi     = {10.48550/arXiv.2405.03998}
}

@inproceedings{chung2022talebrush,
  author    = {Chung, John Joon Young and Kim, Wooseok and Yoo, Kang Min and Lee, Hwaran and Adar, Eytan and Chang, Minsuk},
  title     = {{TaleBrush}: Sketching Stories with Generative Pretrained Language Models},
  booktitle = CHI,
  pages     = {1--19},
  year      = {2022},
  doi       = {10.1145/3491102.3501819}
}

@article{pan2024smartboard,
  author  = {Liu, Ziao and Xie, Xiao and He, Moqi and Zhao, Wenshuo and Wu, Yihong and Cheng, Liqi and Zhang, Hui and Wu, Yingcai},
  title   = {Smartboard: Visual Exploration of Team Tactics with {LLM} Agent},
  journal = TVCG,
  volume  = {31},
  number  = {1},
  pages   = {23--33},
  year    = {2025},
  doi     = {10.1109/TVCG.2024.3456200}
}

@inproceedings{qian2024shapeit,
  author    = {Qian, Wanli and Gao, Chenfeng and Sathya, Anup and Suzuki, Ryo and Nakagaki, Ken},
  title     = {{SHAPE-IT}: Exploring Text-to-Shape-Display for Generative Shape-Changing Behaviors with {LLMs}},
  booktitle = UIST,
  pages     = {1--29},
  year      = {2024},
  doi       = {10.1145/3654777.3676348}
}

@inproceedings{angert2023spellburst,
  author    = {Angert, Tyler and Suzara, Miroslav and Han, Jenny and Pondoc, Christopher and Subramonyam, Hariharan},
  title     = {Spellburst: A Node-based Interface for Exploratory Creative Coding with Natural Language Prompts},
  booktitle = UIST,
  pages     = {1--22},
  year      = {2023},
  doi       = {10.1145/3586183.3606719}
}

@inproceedings{jiang2025neurosync,
  author    = {Zhang, Wenshuo and Shen, Leixian and Xu, Shuchang and Wang, Jindu and Zhao, Jian and Qu, Huamin and Yuan, Lin-Ping},
  title     = {{NeuroSync}: Intent-Aware Code-Based Problem Solving via Direct {LLM} Understanding Modification},
  booktitle = UIST,
  pages     = {1--19},
  year      = {2025},
  doi       = {10.1145/3746059.3747668}
}

@inproceedings{cai2024coladder,
  author    = {Yen, Ryan and Zhu, Jiawen Stefanie and Suh, Sangho and Xia, Haijun and Zhao, Jian},
  title     = {{CoLadder}: Manipulating Code Generation via Multi-Level Blocks},
  booktitle = UIST,
  year      = {2024},
  pages     = {1--20},
  doi       = {10.1145/3654777.3676357}
}

@article{wang2024formulator,
  author  = {Wang, Chenglong and Thompson, John and Lee, Bongshin},
  title   = {{Data Formulator}: {AI}-Powered Concept-Driven Visualization Authoring},
  journal = TVCG,
  volume  = {30},
  number  = {1},
  pages   = {1128--1138},
  year    = {2024},
  doi     = {10.1109/TVCG.2023.3326585}
}

@inproceedings{wang2025formulator2,
  author    = {Wang, Chenglong and Lee, Bongshin and Drucker, Steven M. and Marshall, Dan and Gao, Jianfeng},
  title     = {{Data Formulator 2}: Iterative Creation of Data Visualizations, with {AI} Transforming Data Along the Way},
  booktitle = CHI,
  year      = {2025},
  pages     = {1--17},
  doi       = {10.1145/3706598.3713296}
}

@inproceedings{ma2023insightpilot,
  author    = {Ma, Pingchuan and Ding, Rui and Wang, Shuai and Han, Shi and Zhang, Dongmei},
  title     = {{InsightPilot}: An {LLM}-Empowered Automated Data Exploration System},
  booktitle = EMNLP,
  year      = {2023},
  pages     = {346--352},
  doi       = {10.18653/v1/2023.emnlp-demo.31}
}

@article{chen2024lightva,
  author  = {Zhao, Yuheng and Wang, Junjie and Xiang, Linbing and Zhang, Xiaowen and Guo, Zifei and Turkay, Cagatay and Zhang, Yu and Chen, Siming},
  title   = {{LightVA}: Lightweight Visual Analytics with {LLM}-Agent-Based Task Planning and Execution},
  journal = TVCG,
  volume  = {31},
  number  = {9},
  pages   = {6162--6177},
  year    = {2025},
  doi     = {10.1109/TVCG.2024.3496112}
}

@article{liu2025interchat,
  author  = {Chen, Juntong and Wu, Jiang and Guo, Jiajing and Mohanty, Vikram and Li, Xueming and Ono, Jorge Piazentin and He, Wenbin and Ren, Liu and Liu, Dongyu},
  title   = {{InterChat}: Enhancing Generative Visual Analytics using Multimodal Interactions},
  journal = CGF,
  volume  = {44},
  number  = {3},
  year    = {2025},
  pages   = {e70112},
  doi     = {10.1111/cgf.70112}
}

@article{zhang2024cocoa,
  author  = {K. J. Kevin Feng and Kevin Pu and Matt Latzke and Tal August and Pao Siangliulue and Jonathan Bragg and Daniel S. Weld and Amy X. Zhang and Joseph Chee Chang},
  title   = {Cocoa: Co-Planning and Co-Execution with {AI} Agents},
  journal = {arXiv preprint arXiv:2412.10999},
  year    = {2024},
  doi     = {10.48550/arXiv.2412.10999}
}

@inproceedings{castelo2025flowco,
  author    = {Freund, Stephen N. and Simon, Brooke and Berger, Emery D. and Jun, Eunice},
  title     = {Flowco: Mixed-Initiative Authoring of Reliable End-to-End Data Analyses via Dataflow Graphs and {LLMs}},
  booktitle = UIST,
  pages     = {1--20},
  year      = {2025},
  doi       = {10.1145/3746059.3747636}
}

@inproceedings{li2025jupybara,
  author    = {Wang, Huichen Will and Birnbaum, Larry and Setlur, Vidya},
  title     = {Jupybara: Operationalizing a Design Space for Actionable Data Analysis and Storytelling with {LLMs}},
  booktitle = CHI,
  pages     = {1--24},
  year      = {2025},
  doi       = {10.1145/3706598.3713913}
}

@article{he2025deepanalyze,
  author  = {Shaolei Zhang and Ju Fan and Meihao Fan and Guoliang Li and Xiaoyong Du},
  title   = {{DeepAnalyze}: Agentic Large Language Models for Autonomous Data Science},
  journal = {arXiv preprint arXiv:2510.16872},
  year    = {2025},
  doi     = {10.48550/arXiv.2510.16872}
}

@article{chen2025proactiveva,
  author  = {Zhao, Yuheng and Shu, Xueli and Fan, Liwen and Gao, Lin and Zhang, Yu and Chen, Siming},
  title   = {{ProactiveVA}: Proactive Visual Analytics with {LLM}-Based {UI} Agent},
  journal = TVCG,
  volume  = {32},
  number  = {1},
  pages   = {451--461},
  year    = {2026},
  doi     = {10.1109/TVCG.2025.3642628}
}

@inproceedings{cheng2025agdebugger,
  author    = {Epperson, Will and Bansal, Gagan and Dibia, Victor C and Fourney, Adam and Gerrits, Jack and Zhu, Erkang (Eric) and Amershi, Saleema},
  title     = {Interactive Debugging and Steering of Multi-Agent {AI} Systems},
  booktitle = CHI,
  year      = {2025},
  pages     = {1--15},
  doi       = {10.1145/3706598.3713581}
}

@inproceedings{hu2024infiagent,
  author  = {Xueyu Hu and Ziyu Zhao and Shuang Wei and Ziwei Chai and Qianli Ma and Guoyin Wang and Xuwu Wang and Jing Su and Jingjing Xu and Ming Zhu and Yao Cheng and Jianbo Yuan and Jiwei Li and Kun Kuang and Yang Yang and Hongxia Yang and Fei Wu},
  title   = {{InfiAgent-DABench}: Evaluating Agents on Data Analysis Tasks},
  booktitle = ICML,
  year    = {2024},
  pages   = {19544--19572}
}

@article{li2024tapilot,
  author  = {Jinyang Li and Nan Huo and Yan Gao and Jiayi Shi and Yingxiu Zhao and Ge Qu and Yurong Wu and Chenhao Ma and Jian-Guang Lou and Reynold Cheng},
  title   = {{Tapilot-Crossing}: Benchmarking and Evolving {LLMs} Towards Interactive Data Analysis Agents},
  journal = {arXiv preprint arXiv:2403.05307},
  year    = {2024},
  doi     = {10.48550/arXiv.2403.05307}
}

@inproceedings{siddiqui2025insightbench,
  author    = {Gaurav Sahu and Abhay Puri and Juan A. Rodriguez and Amirhossein Abaskohi and Mohammad Chegini and Alexandre Drouin and Perouz Taslakian and Valentina Zantedeschi and Alexandre Lacoste and David Vazquez and Nicolas Chapados and Christopher Pal and Sai Rajeswar and Issam H. Laradji},
  title     = {{InsightBench}: Evaluating Business Analytics Agents Through Multi-Step Insight Generation},
  booktitle = ICLR,
  year      = {2025}
}

@article{zhang2025condabench,
  author  = {Avik Dutta and Priyanshu Gupta and Hosein Hasanbeig and Rahul Pratap Singh and Harshit Nigam and Sumit Gulwani and Arjun Radhakrishna and Gustavo Soares and Ashish Tiwari},
  title   = {{ConDABench}: Interactive Evaluation of Language Models for Data Analysis},
  journal = {arXiv preprint arXiv:2510.13835},
  year    = {2025},
  doi     = {10.48550/arXiv.2510.13835}
}

@article{zhang2025dacomp,
  author  = {Fangyu Lei and Jinxiang Meng and Yiming Huang and Junjie Zhao and Yitong Zhang and Jianwen Luo and Xin Zou and Ruiyi Yang and Wenbo Shi and Yan Gao and Shizhu He and Zuo Wang and Qian Liu and Yang Wang and Ke Wang and Jun Zhao and Kang Liu},
  title   = {{DAComp}: Benchmarking Data Agents across the Full Data Intelligence Lifecycle},
  journal = {arXiv preprint arXiv:2512.04324},
  year    = {2025},
  doi     = {10.48550/arXiv.2512.04324}
}

@article{zhang2026dsaeval,
  author  = {Maojun Sun and Yifei Xie and Yue Wu and Ruijian Han and Binyan Jiang and Defeng Sun and Yancheng Yuan and Jian Huang},
  title   = {{DSAEval}: Evaluating Data Science Agents on a Wide Range of Real-World Data Science Problems},
  journal = {arXiv preprint arXiv:2601.13591},
  year    = {2026},
  doi     = {10.48550/arXiv.2601.13591}
}

@article{rao2025noteveryone,
  author  = {Ma, Qianou and Koedinger, Kenneth and Wu, Tongshuang},
  title   = {Not Everyone Wins with {LLMs}: Behavioral Patterns and Pedagogical Implications for {AI} Literacy in Programmatic Data Science},
  journal = {arXiv preprint arXiv:2509.21890},
  year    = {2025},
  doi     = {10.48550/arXiv.2509.21890}
}

@inproceedings{kang2025guardrails,
  author    = {Ding, Zijian and Brachman, Michelle and Chan, Joel and Geyer, Werner},
  title     = {``{The} Diagram is like Guardrails'': Structuring {GenAI}-Assisted Hypotheses Exploration with an Interactive Shared Representation},
  booktitle = CANDC,
  year      = {2025},
  pages     = {606--625},
  doi       = {10.1145/3698061.3726935}
}

@article{zhao2025insightlens,
  author  = {Weng, Luoxuan and Wang, Xingbo and Lu, Junyu and Feng, Yingchaojie and Liu, Yihan and Feng, Haozhe and Huang, Danqing and Chen, Wei},
  title   = {{InsightLens}: Augmenting {LLM}-Powered Data Analysis With Interactive Insight Management and Navigation},
  journal = TVCG,
  year    = {2025},
  volume  = {31},
  number  = {6},
  pages   = {3719--3732},
  doi     = {10.1109/TVCG.2025.3567131}
}

@inproceedings{liu2024waitgpt,
  author    = {Xie, Liwenhan and Zheng, Chengbo and Xia, Haijun and Qu, Huamin and Zhu-Tian, Chen},
  title     = {{WaitGPT}: Monitoring and Steering Conversational {LLM} Agent in Data Analysis with On-the-Fly Code Visualization},
  booktitle = UIST,
  year      = {2024},
  pages     = {1--14},
  doi       = {10.1145/3654777.3676374}
}

@inproceedings{jiang2023graphologue,
  author    = {Jiang, Peiling and Rayan, Jude and Dow, Steven P. and Xia, Haijun},
  title     = {Graphologue: Exploring Large Language Model Responses with Interactive Diagrams},
  booktitle = UIST,
  year      = {2023},
  pages     = {1--20},
  doi       = {10.1145/3586183.3606737}
}

@inproceedings{suh2023sensecape,
  author    = {Suh, Sangho and Min, Bryan and Palani, Srishti and Xia, Haijun},
  title     = {Sensecape: Enabling Multilevel Exploration and Sensemaking with Large Language Models},
  booktitle = UIST,
  year      = {2023},
  pages     = {1--18},
  doi       = {10.1145/3586183.3606756}
}

@inproceedings{suh2024luminate,
  author    = {Suh, Sangho and Chen, Meng and Min, Bryan and Li, Toby Jia-Jun and Xia, Haijun},
  title     = {Luminate: Structured Generation and Exploration of Design Space with Large Language Models for Human-{AI} Co-Creation},
  booktitle = CHI,
  year      = {2024},
  pages     = {1--26},
  doi       = {10.1145/3613904.3642400}
}

@article{wu2024intelligentcanvas,
  author  = {Zijian Ding and Joel Chan},
  title   = {Intelligent Canvas: Enabling Design-Like Exploratory Visual Data Analysis with Generative {AI} through Rapid Prototyping, Iteration and Curation},
  journal = {arXiv preprint arXiv:2402.08812},
  year    = {2024},
  doi     = {10.48550/arXiv.2402.08812}
}

@inproceedings{lee2023memorysandbox,
  author    = {Huang, Ziheng and Gutierrez, Sebastian and Kamana, Hemanth and Macneil, Stephen},
  title     = {{Memory Sandbox}: Transparent and Interactive Memory Management for Conversational Agents},
  booktitle = UISTADJ,
  year      = {2023},
  pages     = {1--3},
  doi       = {10.1145/3586182.3615796}
}

@inproceedings{kim2024hallmark,
  author    = {Hoque, Md Naimul and Mashiat, Tasfia and Ghai, Bhavya and Shelton, Cecilia D. and Chevalier, Fanny and Kraus, Kari and Elmqvist, Niklas},
  title     = {The {HaLLMark} Effect: Supporting Provenance and Transparent Use of Large Language Models in Writing with Interactive Visualization},
  booktitle = CHI,
  year      = {2024},
  pages     = {1--15},
  doi       = {10.1145/3613904.3641895}
}

@inproceedings{gu2024steering,
  author    = {Kazemitabaar, Majeed and Williams, Jack and Drosos, Ian and Grossman, Tovi and Henley, Austin Zachary and Negreanu, Carina and Sarkar, Advait},
  title     = {Improving Steering and Verification in {AI}-Assisted Data Analysis with Interactive Task Decomposition},
  booktitle = UIST,
  year      = {2024},
  publisher = {1--19},
  doi       = {10.1145/3654777.3676345}
}

@article{ragan2015provenance,
  author  = {Ragan, Eric D. and Endert, Alex and Sanyal, Jibonananda and Chen, Jian},
  title   = {Characterizing Provenance in Visualization and Data Analysis: An Organizational Framework of Provenance Types and Purposes},
  journal = TVCG,
  volume  = {22},
  number  = {1},
  pages   = {31--40},
  year    = {2016},
  doi     = {10.1109/TVCG.2015.2467551}
}

@article{xu2019provenance,
  author  = {Madanagopal, Karthic and Ragan, Eric D. and Benjamin, Perakath},
  title   = {Analytic Provenance in Practice: The Role of Provenance in Real-World Visualization and Data Analysis Environments},
  journal = CGANDA,
  volume  = {39},
  number  = {6},
  pages   = {30--45},
  year    = {2019},
  doi     = {10.1109/MCG.2019.2933419}
}

@inproceedings{deutch2021diy,
  author    = {Narechania, Arpit and Fourney, Adam and Lee, Bongshin and Ramos, Gonzalo},
  title     = {{DIY}: Assessing the Correctness of Natural Language to {SQL} Systems},
  booktitle = IUI,
  year      = {2021},
  pages     = {597--607},
  doi       = {10.1145/3397481.3450667}
}

@article{epperson2023autoprofiler,
  author  = {Epperson, Will and Gorantla, Vaishnavi and Moritz, Dominik and Perer, Adam},
  title   = {Dead or Alive: Continuous Data Profiling for Interactive Data Science},
  journal = TVCG,
  year    = {2024},
  volume  = {30},
  number  = {1},
  pages   = {197--207},
  doi     = {10.1109/TVCG.2023.3327367}
}

@article{epperson2025noteflow,
  author    = {Yuan Tian and Dazhen Deng and Sen Yang and Huawei Zheng and Bowen Shi and Kai Xiong and Xinjing Yi and Yingcai Wu},
  title     = {{NoteFlow}: Recommending Charts as Sight Glasses for Tracing Data Flow in Computational Notebooks},
  journal   = {arXiv preprint arXiv:2502.02326},
  year      = {2025},
  doi       = {10.48550/arXiv.2502.02326}
}

@inproceedings{pu2021datamations,
  author    = {Pu, Xiaoying and Kross, Sean and Hofman, Jake M. and Goldstein, Daniel G.},
  title     = {Datamations: Animated Explanations of Data Analysis Pipelines},
  booktitle = CHI,
  year      = {2021},
  pages     = {1--14},
  doi       = {10.1145/3411764.3445063}
}

@article{nguyen2016sensepath,
  author  = {Nguyen, Phong H. and Xu, Kai and Wheat, Ashley and Wong, B. L. William and Attfield, Simon and Fields, Bob},
  title   = {{SensePath}: Understanding the Sensemaking Process Through Analytic Provenance},
  journal = TVCG,
  volume  = {22},
  number  = {1},
  pages   = {41--50},
  year    = {2016},
  doi     = {10.1109/TVCG.2015.2467611}
}

@inproceedings{srinivasan2021snowy,
  author    = {Srinivasan, Arjun and Setlur, Vidya},
  title     = {Snowy: Recommending Utterances for Conversational Visual Analysis},
  booktitle = UIST,
  year      = {2021},
  pages     = {864--880},
  doi       = {10.1145/3472749.3474792}
}

@inproceedings{head2019foraging,
  author    = {Kery, Mary Beth and John, Bonnie E. and O'Flaherty, Patrick and Horvath, Amber and Myers, Brad A.},
  title     = {Towards Effective Foraging by Data Scientists to Find Past Analysis Choices},
  booktitle = CHI,
  year      = {2019},
  pages     = {1--13},
  doi       = {10.1145/3290605.3300322}
}

@inproceedings{wang2025inreactable,
  author    = {Aodeng, Gerile and Li, Guozheng and Feng, Yunshan and Chen, Qiyang and Zhang, Yu and Liu, Chi Harold},
  title     = {{InReAcTable}: {LLM}-powered Interactive Visual Data Story Construction from Tabular Data},
  booktitle = UIST,
  year      = {2025},
  pages     = {1--16},
  doi       = {10.1145/3746059.3747719}
}

@inproceedings{he2024datanarrative,
  author    = {Islam, Mohammed Saidul and Laskar, Md Tahmid Rahman  and Parvez, Md Rizwan and Hoque, Enamul and Joty, Shafiq},
  title     = {{DataNarrative}: Automated Data-Driven Storytelling with Visualizations and Texts},
  booktitle = EMNLP,
  year      = {2024},
  pages     = {19253--19286},
  doi       = {10.18653/v1/2024.emnlp-main.1073}
}

@inproceedings{epperson2025respark,
  author    = {Tian, Yuan and Zhang, Chuhan and Wang, Xiaotong and Pan, Sitong and Cui, Weiwei and Zhang, Haidong and Deng, Dazhen and Wu, Yingcai},
  title     = {{ReSpark}: Leveraging Previous Data Reports as References to Generate New Reports with {LLMs}},
  booktitle = UIST,
  year      = {2025},
  pages     = {1--18},
  doi       = {10.1145/3746059.3747644}
}

@article{sultanum2026narrative,
  author  = {Oliver Huang and Muhammad Fatir and Steven Luo and Sangho Suh and Hariharan Subramonyam and Carolina Nobre},
  title   = {Narrative Scaffolding: A Narrative-First Framework for Data-Driven Sensemaking},
  journal = {arXiv preprint arXiv:2512.18920},
  year    = {2025},
  doi     = {10.48550/arXiv.2512.18920}
}

@article{gu2025magneticui,
  author  = {Hussein Mozannar and Gagan Bansal and Cheng Tan and Adam Fourney and Victor Dibia and Jingya Chen and Jack Gerrits and Tyler Payne and Matheus Kunzler Maldaner and Madeleine Grunde-McLaughlin and Eric Zhu and Griffin Bassman and Jacob Alber and Peter Chang and Ricky Loynd and Friederike Niedtner and Ece Kamar and Maya Murad and Rafah Hosn and Saleema Amershi},
  title   = {{Magentic-UI}: Towards Human-in-the-Loop Agentic Systems},
  journal = {arXiv preprint arXiv:2507.22358},
  year    = {2025},
  doi     = {10.48550/arXiv.2507.22358}
}

@article{gathani2024provenance,
  author  = {Shaghayegh Esmaeili and Irelis D. Suarez and Ezekiel Ajayi and Eric D. Ragan},
  title   = {Empirical Insights into Analytic Provenance Summarization: A Study on Segmenting Data Analysis Workflows},
  journal = {arXiv preprint arXiv:2410.11011},
  year    = {2024},
  doi     = {10.48550/arXiv.2410.11011}
}

@inproceedings{chen2025visegpt,
  author    = {Zhu, Jiajun and Cheng, Xinyu and Luo, Zhongsu and Zhou, Yunfan and Shu, Xinhuan and Weng, Di and Wu, Yingcai},
  title     = {{ViseGPT}: Towards Better Alignment of {LLM}-generated Data Wrangling Scripts and User Prompts},
  booktitle = UIST,
  year      = {2025},
  pages     = {1--16},
  doi       = {10.1145/3746059.3747689}
}

@article{wu2025vispilot,
  author  = {Zhen Wen and Luoxuan Weng and Yinghao Tang and Runjin Zhang and Yuxin Liu and Bo Pan and Minfeng Zhu and Wei Chen},
  title   = {Exploring Multimodal Prompt for Visualization Authoring with Large Language Models},
  journal = {arXiv preprint arXiv:2504.13700},
  year    = {2025},
  doi     = {10.48550/arXiv.2504.13700}
}

@inproceedings{manatkar2024quis,
  author    = {Manatkar, Abhijit and Akella, Ashlesha and Gupta, Parthivi and Narayanam, Krishnasuri},
  title     = {{QUIS}: Question-Guided Insights Generation for Automated Exploratory Data Analysis},
  booktitle = EMNLP,
  year      = {2024},
  pages     = {1523--1535},
  doi       = {10.18653/v1/2024.emnlp-industry.111}
}

@inproceedings{hong2025datainterpreter,
  author    = {Sirui Hong and Yizhang Lin and Bang Liu and Bangbang Liu and Binhao Wu and Ceyao Zhang and Danyang Li and Jiaqi Chen and Jiayi Zhang and Jinlin Wang and Li Zhang and Lingyao Zhang and Min Yang and Mingchen Zhuge and Taicheng Guo and Tuo Zhou and Wei Tao and Robert Tang and Xiangtao Lu and Xiawu Zheng and Xinbing Liang and Yaying Fei and Yuheng Cheng and Yongxin Ni and Zhibin Gou and Zongze Xu and Yuyu Luo and Chenglin Wu},
  title     = {{Data Interpreter}: An {LLM} Agent for Data Science},
  booktitle = FINDACL,
  year      = {2025},
  pages     = {19796--19821},
  doi       = {10.18653/v1/2025.findings-acl.1016}
}

@inproceedings{majumder24datavoyaer,
  author    = {Bodhisattwa Prasad Majumder and Harshit Surana and Dhruv Agarwal and Sanchaita Hazra and Ashish Sabharwal and Peter Clark},
  title     = {Position: Data-driven Discovery with Large Generative Models},
  booktitle = ICML,
  pages     = {34350--34382},
  year      = {2024},
}

@inproceedings{yao2023react,
  title     = {{ReAct}: Synergizing Reasoning and Acting in Language Models},
  author    = {Shunyu Yao and Jeffrey Zhao and Dian Yu and Nan Du and Izhak Shafran and Karthik R Narasimhan and Yuan Cao},
  booktitle = ICLR,
  year      = {2023}
}

@article{tanahashi2012design,
  title   = {Design Considerations for Optimizing Storyline Visualizations},
  author  = {Tanahashi, Yuzuru and Ma, Kwan-Liu},
  journal = TVCG,
  volume  = {18},
  number  = {12},
  pages   = {2679--2688},
  year    = {2012},
  doi     = {10.1109/TVCG.2012.212}
}

@article{tang2019istoryline,
  author   = {Tang, Tan and Rubab, Sadia and Lai, Jiewen and Cui, Weiwei and Yu, Lingyun and Wu, Yingcai},
  journal  = TVCG,
  title    = {{iStoryline}: Effective Convergence to Hand-drawn Storylines},
  year     = {2019},
  volume   = {25},
  number   = {1},
  pages    = {769--778},
  doi      = {10.1109/TVCG.2018.2864899}
}

@article{wang2020datashot,
  author   = {Wang, Yun and Sun, Zhida and Zhang, Haidong and Cui, Weiwei and Xu, Ke and Ma, Xiaojuan and Zhang, Dongmei},
  journal  = TVCG,
  title    = {DataShot: Automatic Generation of Fact Sheets from Tabular Data},
  year     = {2020},
  volume   = {26},
  number   = {1},
  pages    = {895--905},
  doi      = {10.1109/TVCG.2019.2934398}
}

@misc{measuringu_sus,
  key          = {{MeasuringU}},
  author       = {Jeff Sauro},
  title        = {Measuring Usability with the System Usability Scale ({SUS})},
  howpublished = {{MeasuringU}},
  month        = feb,
  year         = {2011},
  note         = {Available: \url{https://measuringu.com/sus} } # LACC
}

@inproceedings{lewis2009factorsus,
  title     = {The Factor Structure of the System Usability Scale},
  author    = {Lewis, James R. and Sauro, Jeff},
  booktitle = HCD,
  pages     = {94--103},
  year      = {2009},
  doi       = {10.1007/978-3-642-02806-9_12},
}

@misc{student_performance_320,
  author       = {Cortez, Paulo},
  title        = {{Student Performance}},
  howpublished = {{UC Irvine} Machine Learning Repository},
  note         = {Available: \url{https://archive.ics.uci.edu/dataset/320/student+performance} } # LACC,
  doi          = {10.24432/C5TG7T}
}

@misc{smith2022videogamesales,
  author    = {Gregory Smith},
  title     = {{Video Games Sales}},
  howpublished = {Zenodo},
  note      = {Available: \url{https://zenodo.org/records/5898311} } # LACC,
  doi       = {10.5281/zenodo.5898311}
}

@incollection{JBrooke1996sus,
  title     = {{SUS}: A ``Quick and Dirty'' Usability Scale},
  author    = {John Brooke},
  booktitle = {Usability Evaluation in Industry},
  pages     = {189--194},
  year      = {1996},
  publisher = {Taylor \& Francis}
}

@article{liu2013storyflow,
  author={Liu, Shixia and Wu, Yingcai and Wei, Enxun and Liu, Mengchen and Liu, Yang},
  journal=TVCG, 
  title={StoryFlow: Tracking the Evolution of Stories}, 
  year={2013},
  volume={19},
  number={12},
  pages={2436--2445},
  doi={10.1109/TVCG.2013.196}}

@article{DBLP:journals/tvcg/SedlmairMM12,
  author       = {Michael Sedlmair and
                  Miriah D. Meyer and
                  Tamara Munzner},
  title        = {Design Study Methodology: Reflections from the Trenches and the Stacks},
  journal      = TVCG, 
  volume       = {18},
  number       = {12},
  pages        = {2431--2440},
  year         = {2012},
  doi          = {10.1109/TVCG.2012.213}
}

\vfill

\end{document}